\pdfoutput=1
\documentclass{article}

\usepackage[main, final]{neurips_2026}
\usepackage[utf8]{inputenc} % allow utf-8 input
\usepackage[T1]{fontenc}    % use 8-bit T1 fonts
\usepackage[hidelinks]{hyperref}
\usepackage{url}            % simple URL typesetting
\usepackage{booktabs}       % professional-quality tables
\usepackage{amsfonts}       % blackboard math symbols
\usepackage{nicefrac}       % compact symbols for 1/2, etc.
\usepackage{microtype}      % microtypography
\usepackage{xcolor}         % colors
\usepackage{xspace}
\usepackage{enumitem}
\usepackage{amsmath}
\usepackage{cleveref}
\usepackage{subcaption}
\usepackage{graphicx}
\usepackage{mdframed}
\usepackage{multirow}
\usepackage{colortbl}
\usepackage{fvextra}
\usepackage[hang,flushmargin]{footmisc}
\usepackage{tcolorbox}
\usepackage{tabularx}

\newcolumntype{T}{>{\hsize=1.45\hsize\raggedright\arraybackslash}X}
\newcolumntype{D}{>{\raggedright\arraybackslash}p{75pt}}

\newcommand{\ignore}[1]{}
\newcommand{\mytt}[1]{\texttt{\small #1}}

\newcommand{\ourmodel}[0]{Gaggle\xspace}

\newcommand{\ourptmodeltt}[0]{\mytt{gaggle-\allowbreak{}nanochat-\allowbreak{}pretrained-\allowbreak{}climbmix-\allowbreak{}20260924}\xspace}

\newcommand{\ourbasereranker}[0]{Gaggle (base reranker)\xspace}
\newcommand{\ourbasererankertt}[0]{\mytt{gaggle-\allowbreak{}reranker-\allowbreak{}20261005}\xspace}

\newcommand{\ourbidirptmodeltt}{\mytt{gaggle-\allowbreak{}nanochat-\allowbreak{}pretrained-\allowbreak{}climbmix-\allowbreak{}bidirectional-\allowbreak{}20260929}\xspace}

\newcommand{\tableformat}{%
  \fontsize{8}{9.5}\selectfont
  \setlength{\tabcolsep}{2pt}%
  \renewcommand{\arraystretch}{1.08}%
}
\title{Project Greenhouse: Progress Toward Fully Open and Sovereign Agentic Search}

\author{Jimmy Lin \quad Sahel Sharifymoghaddam \quad Lingwei Gu \quad Nour Jedidi\\[1ex]
David R. Cheriton School of Computer Science\\
University of Waterloo
}

\begin{document}

\maketitle

\begin{abstract}
Project Greenhouse represents our exploration of a simple thesis:
We believe that it is possible to build fully open and sovereign models for agentic search with only modest computational resources.
As a first milestone, we describe how to build a competitive pointwise decoder-only reranker using a simple two-step recipe comprising pre-training from scratch followed by supervised fine-tuning, starting only from commonly available datasets.
Contrary to the dominant approach in the literature, we do not rely on existing open-weight backbones from third parties, and thus we are fully in control of model training, from end to end.
We were able to accomplish the bulk of our experiments using no more than a handful of GPUs.
This report articulates the importance and benefits of our approach, and we share artifacts that enable transparent, independent reproduction of all aspects of model training.
Beyond data, code, and configurations that capture our efforts, we also release checkpoints for our family of \ourmodel models, demonstrating the feasibility of our approach and providing a first step toward validating our broader thesis.
\end{abstract}

\section{Introduction}
\label{section:intro}

Project Greenhouse represents our exploration of a simple thesis:

\begin{quote}
We believe that it is possible to build fully open and sovereign models for agentic search with only modest computational resources.
\end{quote}

This report describes our first project milestone.
As the headline result, we show that it is possible to build a competitive pointwise decoder-only reranker using a simple two-step recipe comprising pre-training from scratch followed by supervised fine-tuning, starting only from commonly available datasets.
Most notably, we do not rely on existing open-weight backbones to initialize our reranker, which has been the dominant approach described in the literature.
Models that we pre-trained ourselves from publicly available raw corpora appear to be sufficient to produce competitive rerankers after supervised fine-tuning using a publicly available dataset.
Our training recipe currently does not require synthetic data or teacher-provided knowledge from any third-party sources (proprietary or open-weight).
Thus, we retain full control over all dependencies in the entire training pipeline, from end to end.
Furthermore, we were able to accomplish the bulk of our experiments using only a handful of GPUs, making it feasible for others to reproduce our results.

In addition to code, data, configurations, and other artifacts necessary to reproduce our results, we also release \ourmodel, a family of fully open and sovereign model checkpoints that encapsulate our efforts.
These artifacts and the experiences we share here offer an ``existence proof'' of one initial important step in our thesis, laying the foundation for future explorations.

To better circumscribe and contextualize our claims, let us be more precise:\

\begin{itemize}

\item Our definition of ``agentic search'' is discussed in Section~\ref{section:agentic-search}.
We are explicit about scoping our efforts around this limited goal, as opposed to training models that are capable of general reasoning, coding, etc.

\item Our definition of ``fully open and sovereign'' is discussed in Section~\ref{section:sovereign}, but intuitively, we desire that {\it every} step of the model training process (data, code, configurations)\ is captured in publicly available artifacts.
 
\end{itemize}

The remainder of this report describes exactly what we've accomplished, starting with the headline result (Section~\ref{section:headline}) and continuing with the details of our pre-training and supervised fine-tuning steps (Sections~\ref{section:pt} and \ref{section:sft}).
Contrastive conditions and variants of our main approach are presented in Section~\ref{section:constrastive-conditions}, before we conclude with a discussion of related work and limitations.

Project Greenhouse is intended to be a long-running project, and this report captures a concrete first milestone that we deem worthwhile to share with the community.
We view the present contributions of this work as follows:

\begin{itemize}

\item We provide a model of LLM training as a directed property hypergraph that allows us to precisely characterize what we mean by ``fully open and sovereign''.
This formulation not only describes our efforts, but can be extended to characterize an open ecosystem and to coordinate distributed efforts by research groups beyond our institution.

\item We describe a small step toward supporting the veracity of our thesis.
Although our ambitions encompass agentic search in its entirety, a competitive reranker represents an important stepping stone.
This report shares our experiences and interesting findings along the way.

\item We share our \ourmodel family of decoder-only pointwise rerankers and model checkpoints that capture important steps in our pre-training and fine-tuning journey.

\end{itemize}

\section{Toward Agentic Search}
\label{section:agentic-search}

Given our thesis, it makes sense to be more precise about what we mean by ``agentic search''.
Stripping away unhelpful anthropomorphization and deliberate attempts at making things sound more complicated than they actually are, agents are simply models using tools in loops.
Anthropic engineers characterize agents as ``LLMs using tools based on environmental feedback in a loop''.\footnote{\url{https://www.anthropic.com/engineering/building-effective-agents}}
In modern parlance, the ``thing'' that does the ``looping'' is typically referred to as the harness.
The harness invokes the model, executes tool calls, returns their results, and maintains the context needed to continue (as well as participating in the decision to stop when the user task is complete).

\subsection{Scoping Our Efforts}

Thus, agentic search is simply agents doing ``searchy stuff''. 
For these applications, models use ``search tools'' in loops, broadly defined to include, for example, grep, BM25, dense vector search, external APIs that perform live web search, etc.
Of course, agentic search can involve non-search tools as well, for example, to read a retrieved document, to execute an analysis, or to capture intermediate findings.
Typically, a model would decompose a complex information need into smaller components, and for each, the model might issue a query, examine retrieved results, identify missing information, and follow up with reformulated queries.
In many cases, the model is also responsible for synthesizing relevant information into a coherent whole that addresses the user's information need.
Our definition is intentionally broad to encompass the information-seeking aspects of a variety of downstream tasks, from automated data science to writing detailed research reports.

Since agentic search covers quite a bit of ground, we are cognizant of the need to appropriately scope our efforts.
We focus only on information seeking:\ finding, assessing, and synthesizing information from a variety of (possibly heterogeneous) sources in response to a user request.
To be clear, we are not attempting to build a general-purpose frontier model; 
for example, coding and general reasoning are outside our immediate scope.
It would be unreasonable to organize our research around competing directly with frontier labs across the full range of capabilities they pursue.
Instead, we believe that a deliberately scoped effort can yield useful, measurable advances while contributing to knowledge in a rigorous fashion.
For us, the focus is on agentic search using fully open and sovereign models.

Our conception of agentic search does not prescribe a particular system architecture.
An information-seeking task might be handled by a single (powerful) model driven by a simple harness, or by multiple agents that explore different sources in parallel and coordinate autonomously to exchange findings.
A tool call can be as simple as a system command (e.g., \mytt{grep}), a REST API invocation, or may itself be a composite system involving LLMs (e.g., BM25 search followed by LLM-based reranking). 
These alternatives represent design choices to be explored empirically.

While our ultimate goal is to build fully open and sovereign models for agentic search in its entirety, our approach emphasizes incremental progress:\ building useful components, rigorously understanding their behavior and limitations under controlled conditions, and composing them into increasingly capable systems.
Achieving full openness and sovereignty while maintaining output quality is an unrealistic goal at the outset:\ instead, initial systems are likely to be hybrids that integrate both proprietary and open components---for example, a proprietary frontier model coordinator that synthesizes relevant information while delegating searches to fully open and sovereign models.
Consistent with our core thesis, we seek to make individual components fully open and sovereign so that others can inspect their development, reproduce our experiments, and build on the results.
Over time, we hope to gradually expand our scope to encompass more and more of agentic search.

\subsection{Pointwise Reranking}
\label{section:agentic-search:pointwise-reranking}

Against this backdrop, our first project milestone---that we have successfully accomplished, as described in this report---focuses on a decoder-only pointwise reranker.
While rerankers are quite common today, let us be precise about the problem we are tackling, building on the definition of text retrieval (ranking) from \cite{Lin_arXiv2021_repir}:

\begin{quote}
The formulation of text retrieval (alternatively, text ranking)---what information retrieval researchers more precisely call {\it ad hoc} retrieval---is typically defined as follows:\
Given an information need expressed as a query $q$, the text retrieval task is to return a ranked list of $k$ documents\footnote{Consistent with parlance in information retrieval, ``document'' is used in a generic sense to refer to the unit of retrieved text, even though in truth it may be a passage, a web page, a PDF, or some arbitrary span of text.} $\{d_1, d_2 \ldots d_k\}$ from an arbitrarily large but finite collection of documents $\mathcal{D} = \{ d_i \}$ that maximizes a metric of interest, for example, nDCG, AP, etc.
These metrics vary, but they all aim to quantify the ``goodness'' of the results with respect to the information need; in some cases, metrics can be understood more formally in terms of the utility that a user would derive from consuming the results.
The retrieval task is also called top-$k$ retrieval (or ranking), where $k$ is the length of the ranked list (also known as the retrieval or ranking depth).
\end{quote}

Given this setup, a reranker is a component that takes as input an initial list of documents (typically from top-$k$ retrieval, often called the first stage) and reorders the list (i.e., generates a permutation of the input) with the aim to improve its quality in terms of the metric of interest.
A {\it pointwise} reranker accomplishes this one $(q, d_n)$ pair at a time, where $q$ is the query and $d_n$ is a document returned by top-$k$ retrieval above, $n \in \{ 1 \ldots k \}$.
For every such pair, which is considered in isolation independently, the reranker generates a score $s_n$, i.e., $\mathcal{R}(q, d_n) = s_n$, such that sorting by these scores produces a permutation of the input ranked list that is intended to increase the metric of interest.\footnote{Note that {\it pointwise} rerankers are often contrasted with {\it pairwise} rerankers that consume {\it pairs} of documents and {\it listwise} rankers that consume {\it lists} of documents. Our initial project milestone focuses on {\it pointwise} reranking, but we hope to build other types of rerankers as part of future work. See additional discussion in Section~\ref{section:agentic-search:pointwise}.}

In this report, we describe a fully open and sovereign decoder-only LLM for pointwise reranking that achieves a level of effectiveness competitive with alternatives reported in the literature built on open-weight models.
Beyond model checkpoints, we share all artifacts (data, code, configurations) that are necessary to reproduce our work, end to end.

\subsection{Why Pointwise Reranking?}
\label{section:agentic-search:pointwise}

A natural next question:\ Why start with a pointwise reranker as the first project milestone?
We provide three answers:

\paragraph{Ease of iteration.}
Velocity is an important consideration in our project, and a pointwise reranker enables rapid iteration during model development.
At inference time, the reranker is simply a function that maps a query--document pair $(q, d)$ to a scalar score.
Each candidate can be scored independently, and the resulting scores determine the ranking.
This gives us a simple interface for training and evaluation, allowing us to isolate the effects of changes to training data, training recipes, model architectures, configurations, etc.

With pointwise rerankers, we can evaluate model checkpoints on a fixed set of candidate documents without processing the entire corpus each time.
Compared to pairwise~\citep{pradeep2021expando} and listwise~\citep{pradeep2023rankvicuna,pradeep2023rankzephyr} variants, pointwise rerankers are easier to build because there are fewer design choices to explore.
With respect to other reasonable options as a first project milestone, we considered but did not pursue training embedding models, which typically involve much more effort.
Assessing the quality of an updated model usually requires recomputing embeddings for the entire corpus before evaluating top-$k$ retrieval.
For large collections, this additional step can substantially slow the development cycle.
We feel that training agentic search models is even less appropriate as a first step, as such efforts involve substantial complexity, including generating and evaluating multi-step trajectories and, depending on the approach, collecting rollouts to feed reinforcement learning.
In the end, we targeted pointwise reranking, which gives us a more tightly scoped experimental setting with fewer interacting components.
We believe that it is a simple but useful capability that can provide an interesting existence proof for our core thesis.

\paragraph{Usefulness in isolation.}
A reranker is useful as a standalone component in a broader retrieval pipeline.
It can be applied to candidate documents generated by any number of approaches:\ BM25, a dense retrieval model, an external search API, or a hybrid of these.
Reranking can be especially helpful in cases where one does not have direct access to the raw corpus, such as when searching via a web API.
In all cases, the interface remains the same:\ given a query and a list of candidate documents, score each pair and reorder the candidates.
The design makes it possible to incorporate a reranker into an existing system with minimal effort.

The modularity of rerankers is particularly attractive for agentic search.
At each iteration, a search tool may return more content than the agent can reasonably inspect.
A reranker can help prioritize which results to read or include in the model's context.
This means that a reranker can be inserted into an agentic search application at multiple points, thus increasing its applicability.

The existence of commercial reranking services provides further evidence that this is a useful capability in its own right.
Examples include Microsoft's semantic ranker in Azure AI Search,\footnote{\url{https://learn.microsoft.com/en-us/azure/search/semantic-search-overview}}
Google Cloud's Ranking API,\footnote{\url{https://docs.cloud.google.com/generative-ai-app-builder/docs/ranking}}
NVIDIA's NeMo Retriever Reranking NIM,\footnote{\url{https://docs.nvidia.com/nim/nemo-retriever/reranking/latest/overview.html}}
and dedicated APIs from Cohere,\footnote{\url{https://docs.cohere.com/v2/docs/rerank}}
Voyage AI,\footnote{\url{https://docs.voyageai.com/docs/reranker}}
and Jina AI.\footnote{\url{https://jina.ai/en-US/reranker/}}
These offerings span hosted APIs, integrated search services, and deployable inference microservices, illustrating the utility of reranking across different retrieval architectures.
The recent release of the Jev ``decision model''\footnote{\url{https://typesafe.ai/blog/introducing-system-one-models-and-jev}} further confirms the utility of reranking as an important independent capability:\ although the model was designed for structured outputs in general, reranking has emerged as a popular use case.

\paragraph{Usefulness as a stepping stone.}
A reranker is also an important stepping stone towards more sophisticated models for agentic search.
Beyond ordering results at inference time to prioritize what is fed to the model (as discussed above), a reranker can be used to assess relevance, i.e., used as an LLM judge~\citep{ZhengLianmin_etal_NeurIPS2023,Upadhyay_etal_arXiv2024,Upadhyay_etal_ICTIR2025}.
A pointwise reranker $\mathcal{R}(q, d_n) = s_n$, if properly calibrated, can provide scores that estimate the probability of relevance, i.e., $s_n \approx P(y_n = 1 \mid q, d_n)$, where $y_n$ denotes binary relevance.
For search agents, a good relevance judge might be a helpful capability when inspecting retrieved results, particularly in cases where they only have access to ``weak'' tools such as grep~\citep{LiZhuofeng_etal_arXiv2026_DCI} or BM25 search~\citep{Hsu_etal_arXiv2026_Pi-serini}.

Modern search agents can be trained with reinforcement learning to acquire search policies~\citep{Zhuang_etal_arXiv2025,JinBowen:2503.09516:2025}, and relevance scores can contribute to reward signals by providing feedback on the relevance of the material retrieved by their actions~\citep{mao-etal-2024-rafe,LiYuan:2505.23794:2025}.
Alternatively, a relevance judge could be used as a component of a verifier to assess the quality of rollouts as part of rejection sampling~\citep{Wu:2505.22648:2025,luo-etal-2026-reinforcing}.
There are many possibilities, and it is not difficult to see how a reranker could serve as an important component in agentic search, either at training time or inference time.

\bigskip

To be clear, this report captures our present findings within a larger effort.
Beyond pointwise reranking, we have plans to build models for listwise reranking, query reformulation, relevance feedback, steering search trajectories, etc.\ towards full agentic search.
A reranker is a good first milestone that provides an existence proof of one initial aspect of our broader vision.
Our reranking model is of immediate utility and provides a stepping stone for subsequent research.

\section{Fully Open and Sovereign Models}
\label{section:sovereign}

Next, let us precisely define what we mean by fully open and sovereign models.

\subsection{LLM Training as Hypergraphs}

Abstractly, we can model LLM training as a directed property hypergraph $\mathcal{G} = (V, E)$ where the vertices ($V$) represent artifacts (for example, a raw corpus, training examples, model weights, etc.) and the edges ($E$) represent computations (including training recipes and associated configuration data) that consume upstream artifacts to generate downstream artifacts.
Following standard notation, a (directed) hyperedge $e \in E$ is formally an ordered pair $(T_e, H_e)$, where $T_e, H_e \subseteq V$, commonly written as $T_e \longrightarrow H_e$.
A hyperedge describes a computation, and we can think of an edge traversal as someone running that computation.
Traversing an edge that already exists can be viewed as a reproducibility attempt.

In this directed hypergraph representation, both vertices and hyperedges are annotated with properties in the form of arbitrary key--value pairs.
One simple possibility is that a vertex is annotated with the Hugging Face dataset that captures the artifact and a hyperedge is annotated with a command invocation identified by a commit id in a particular GitHub repo.
Both vertices and edges might be associated with publications.
Each hyperedge might be further annotated with compute requirements, for example, quantifying the approximate number of GPU hours that an edge traversal will require.

For example, pre-training starts from a raw corpus of texts $T_1 = \{\mathcal{C}_1\}$ to produce pre-trained model weights $H_1 = \{\mathcal{M}_1\}$ via $e_1$, a pre-training recipe (Section~\ref{section:pt}).
With appropriate property annotations, it would be possible for someone with sufficient compute resources to reproduce model pre-training, i.e., to traverse $T_1 \longrightarrow H_1$.
In this example, we assume that the pre-training recipe associated with $e_1$ includes not only the actual source code but also all relevant configuration data.

Hyperedges provide a natural representation to capture other aspects of LLM training because they allow multiple tail vertices.
For example, a fine-tuning hyperedge $e_2$ might start from a pre-trained backbone and some training set ($T_2 = \{ \mathcal{M}_1, \mathcal{S}_1 \}$) to produce model weights $H_2 = \{\mathcal{M}_2\}$ via a particular fine-tuning recipe (Section~\ref{section:sft}).
In principle, the head vertex set can contain multiple elements as well, representing more than one output artifact.

\begin{quote}
{\bf Definition:} Given a hypergraph representation of model training, we define a fully open and sovereign model as a vertex (model weights) where {\it all} upstream vertices and edges are publicly available and do not contain proprietary dependencies.
\end{quote}

In other words, a model is fully open and sovereign if and only if it is possible to independently derive the model weights from a set of publicly available sources via computational steps (represented by edge traversals) that are also captured in public artifacts.

Of course, the tracing of upstream sources must ``stop'' somewhere, because otherwise it is easy to arrive at absurdities.
For example, a standard pre-training corpus (e.g., FineWeb-Edu or ClimbMix) lies downstream of potentially complex processes that generated it.
These include stages that involve web crawling, curation, cleaning, parsing, etc., and the texts ultimately derive from a multitude of authors on the web.
It would be impractical to trace LLM training all the way back to these authors.
Similarly, standard training datasets (e.g., MS MARCO) are the output of some annotation process (in this case, by Microsoft).
As we trace sources upstream, where this process ``ends'' necessarily involves making some judgment calls, which we indeed do, based on commonly accepted datasets in the field.
For example, we consider starting with the ClimbMix corpus or the MS MARCO dataset fully open and sovereign, given their widespread availability and familiarity among researchers.

Based on our definition, any model that is derived from open weights, but whose pre-training corpus and pre-training recipes are unavailable, does {\it not} qualify as being fully open and sovereign.
While open-weight models are undoubtedly useful and provide high-quality backbones for subsequent fine-tuning, our work shows that they are not necessary to produce a competitive reranker.
In Section~\ref{section:related-work}, we discuss related work with respect to our definitions.

\subsection{The LLM Training Ecosystem}

The directed hypergraph representation of model training proposed above provides a precise way to capture provenance:\ a model can be characterized as a specific traversal of the training hypergraph, with intermediate artifacts captured along the way.
Such a representation can help a group of collaborating researchers more accurately track ablations and contrastive conditions.
A data ablation can be thought of as a hyperedge with fewer tail vertices.
A code ablation or an alternative training recipe can be thought of as different hyperedges with the same tail vertices.

We can extend this representation of LLM training beyond a single research group to characterize the entire ecosystem comprised of a multitude of researchers around the world working in a distributed fashion.
All that would be needed is a common machine-readable encoding of the representation proposed here and some method of aggregating such representations across many researchers.
With such a characterization of the field, we could, for example, trace all models that are fine-tuned versions of a particular Qwen backbone, or all models that have used MS MARCO or BEIR.
Some of this information is already captured in Hugging Face metadata, but we can imagine an agent semi-automatically enriching the ecosystem description by scanning the latest literature to extract vertex and hyperedge information.

This representation could support agents that identify unexplored combinations of artifacts and recipes, propose new experiments, and execute feasible candidates subject to resource and evaluation constraints.
We could further close the loop with automatic evaluation harnesses and give agents autonomy in ``growing'' the hypergraph according to some high-level goal.
This might provide a concrete substrate for automated AI scientists or for recursive self improvement within a narrowly specified scope.

\subsection{Other Considerations}
\label{section:sovereign:considerations}

\paragraph{Who traverses the edges?}
Viewing model training as traversing hyperedges from tails to heads does not necessarily mean that we have exhaustively traversed every edge {\it ourselves}.
Furthermore, it also does not necessarily mean that a specific hyperedge traversal is guaranteed to be ``successful'' (more discussion below).
It is important to untangle questions about openness and sovereignty (the first) from questions about reproducibility (the second), which are orthogonal.

Openness and sovereignty focus on the {\it availability} of resources (data, code, etc.)\ to transform one artifact into another.
Whether the data and recipes ``work as advertised'' is a (different) question of reproducibility.
They are completely independent.
To see why, consider the hypothetical case in which an organization shares its data and code with another under a non-disclosure agreement.
In this case, the results may be reproducible (by the second organization), but not open since the data and code are not publicly available.

The success of a specific hyperedge traversal is a question of reproducibility, and in this work we make judgment calls.
As a specific example, in one set of experiments we start with a model checkpoint shared by Andrej Karpathy, which he asserts as having been pre-trained on a particular dataset with code in a specific repo.
Although we have {\it not} actually run his code from the dataset, we trust that his complete recipe works as claimed.
That is, we have not independently verified reproducibility, but in this case, we feel that trust is warranted.

Similarly, readers of our report or those wishing to use our results do not necessarily need to traverse every hyperedge themselves.
Instead, they can take advantage of any vertex along a traversal (i.e., directly use the model weights or any intermediate artifacts).
As discussed above, whether a particular code repository accurately reflects the putative hyperedge is a question of reproducibility.
Someone can indeed ``trust us'', for example, that our claimed approach to fine-tuning does indeed yield a particular level of effectiveness on benchmark datasets, or {\it not} trust our claims and try out the method independently (i.e., a reproduction attempt).
This is possible since by our definition every vertex and hyperedge must be publicly available.
Thus, a fully open and sovereign ecosystem is maximally flexible.
Anyone can take advantage of an artifact directly (i.e., download model weights and perform inference) or trace an artifact upstream to rerun the experiments that generated it.
In an open ecosystem, anyone can consider alternatives by branching off another hyperedge and sharing the resulting artifacts, which represents a discrete contribution to our collective knowledge.

\paragraph{What's our standard of reproducibility?}
In the ideal case, repeated traversals of the same hyperedge should yield identical results.
This is a realistic assumption for deterministic transformations.
However, we expect that, in many cases, repeated traversals of the same hyperedge yield ``substantively'' similar, but not necessarily identical, results.

For model training, even with the same inputs, code, configuration, random seed, etc., unavoidable differences in hardware, software environments, and nondeterministic operations in training will likely lead to different model weights (see explorations in Section~\ref{section:sft:reproducibility}).
Thus, our notion of reproducibility does {\it not} require output artifacts to be byte-wise identical.\footnote{As Codex likes to say.}
Strictly speaking, multiple hyperedge traversals are unlikely to produce the same exact output artifact(s); vertices in our LLM training hypergraph thus capture the output of a specific execution instance.

Instead, we adopt a more ``relaxed'' assumption about reproducibility.
Accepting that byte-identical artifacts are unlikely, rerunning a computation should nevertheless reproduce its associated experimental claims substantively, for example, comparable effectiveness on the reported benchmarks, within a reasonable range of variation.
Our use of vague weasel terms such as ``substantive'', ``reasonable'', etc.\ is deliberate, as we will need to exercise judgment, but the community will ultimately assess the veracity of our claims.
Along the way, we will adopt commonly accepted standards, such as quantifying effectiveness variations through multiple trials and referencing existing evaluations already documented in the literature.
Ultimately, the public availability of all artifacts (vertices) and code (hyperedges) makes third-party audits of our claims possible.

\paragraph{What about the use of synthetic data, proprietary teachers, etc.?}
Synthetic data, teacher guidance, and other techniques can significantly increase model effectiveness.
The obvious concern is that they introduce dependencies (both vertices and hyperedges) on proprietary APIs or models, or on models that are themselves not fully open and sovereign.
What do we do?
Here, we attempt to adopt a practical position and exercise judgment on a case-by-case basis---but this is an issue that we frequently grapple with throughout the project.

Consider a concrete case:\
Is the use of synthetic data compatible with our notion of openness and sovereignty?
Let's assume that the dataset is made publicly available, but was generated by a proprietary model.
Let's even assume that the associated code was made available (seeds, prompt templates, etc.)\ but that the data generation process itself requires calling a proprietary API.
Would we consider a model derived from such a dataset fully open and sovereign?
Does the answer to the above question change if the dataset were generated by an open-weight model, in which case it would be possible to recreate the dataset independently?
The same question applies to teacher guidance and other side information, including labels, rankings, scores, and logits.

In short, we are not committing to a specific conclusion in this report, because the actual answer is likely to be organization- and context-dependent.
Nevertheless, we suspect that an important consideration is whether the use of proprietary or non-open models represents a one-time process (e.g., whose results are captured statically in a dataset) or requires continued access during training (e.g., to determine rewards).
However, our commitment to making vertices and hyperedges publicly available allows others to reach their own conclusions about openness and sovereignty.
Furthermore, the transparency of our approach allows others to examine the impact of proprietary and non-open components via ablation studies.

\subsection{Why is this important? The Organizational Perspective}
\label{section:sovereign:organizational}

Sovereign AI encompasses many overlapping and closely related concepts, centered on the idea that organizations---and, by extension, countries---should retain control over their use and governance of AI. 
One aspect of sovereignty is the ability to independently build, deploy, and maintain AI infrastructure without requiring the continued cooperation of an external provider.
It is this aspect of sovereignty, focused specifically on models, that we focus on here.

The ability of organizations to build and deploy their own models enables autonomy and mitigates risks arising from external dependencies.
Reliance on an external model provider exposes an organization to changes in pricing, terms of service, permitted uses, and even continued access.
This is not an abstract concern; recent episodes have raised alarms with many leaders.
On June 12, 2026, Anthropic suspended access to Fable 5 and Mythos 5 for all customers following a US government directive restricting access by foreign nationals, including those within the United States.\footnote{\url{https://www.anthropic.com/news/fable-mythos-access}}
Although these export controls were subsequently lifted,\footnote{\url{https://www.anthropic.com/news/redeploying-fable-5}} the disruption illustrates how organizations can lose access to critical capabilities through decisions outside their control.

Geopolitical risks aside, an organization can be at the mercy of an external provider that determines the availability of certain capabilities for particular applications.
Although such dependencies invariably exist with {\it any} external service, the increasing importance of AI in all aspects of an organization's operations increases the risk from not ``owning the means of production''.
To provide a concrete example:\
At its launch, Anthropic's Fable 5 model routed requests flagged by its safeguards---including requests concerning biology and cybersecurity---to a less capable model.\footnote{\url{https://www.anthropic.com/news/claude-fable-5-mythos-5}}
Regardless of the motivation, such restrictions mean that an organization's access to certain capabilities may depend on the provider's policy, over which the organization may have limited control.

Use of an external model provider also raises questions about control over an organization's proprietary knowledge in the form of prompts, retrieved documents, and other input that may contain confidential information.
\cite{Palantir_2026} advises that an organization should insist that model providers operate under a Zero Data Retention (ZDR) agreement, a privacy concept whereby none of an organization's data in the form of prompts, outputs, and telemetry is retained beyond the ephemeral processing required to fulfill the request. 
However, the report goes on to argue that ZDR may not be sufficient as there are other channels for ``leakage'', and that ZDR does not guarantee continued access, tying in to the other concerns expressed above.

In this context, our notion of fully open and sovereign models gives an organization greater control over all aspects of a model's lifecycle, from end to end. 
Access to model weights alone (i.e., building on existing open-weight models) may not be sufficient to achieve a desired level of control that an organization would be comfortable with.
Developers of open-weight models make checkpoints available, but they do not necessarily disclose the raw pre-training corpora or other important aspects of the training pipeline, including data processing, alignment, the mid- and post-training phases, etc.
While open-weight models certainly provide useful starting points for deployment, they leave important upstream dependencies unresolved.

We identify two main challenges with open-weight models:

First, incomplete provenance makes it difficult to assess whether the model was trained in a way that satisfies an organization's requirements in terms of data licensing, data privacy, data quality, etc.
Furthermore, it can be challenging to assess alignment and whether model behavior is consistent with an organization's values.
For example, it is well known that LLMs can exhibit systematic political and cultural biases associated with their training data and development choices~\citep{feng-etal-2023-pretraining,Buyl:2410.18417:2024}.
For example, a Chinese open-weight model may deny the historic reality of the 1989 Tiananmen Square protests and massacre to reflect the position of the Chinese Communist Party.
More insidious are backdoors that can be embedded in LLMs~\citep{kurita-etal-2020-weight,xu-etal-2024-instructions,Hubinger:2401.05566:2024}.
Both of these issues persist even if an organization performs model inference itself in isolation, eliminating the possibility that the model communicates with the original developers.
Although researchers have developed techniques to ``re-align'' open-weight models~\citep{ChenShengzhuang:2608.27147:2026}, successful realignment on evaluated behaviors does not establish that all unwanted behaviors or hidden backdoors have been removed.
The unknown unknowns remain a concern.

Second, there is no guarantee that the original developers of an open-weight model will continue to release improved checkpoints under the same license.
Future releases might remain proprietary, adopt more restrictive terms, or prioritize capabilities that do not match an organization's needs.
An organization can continue to use an existing checkpoint and improve it through further training, but access to weights alone severely restricts the types of experiments and modifications that can be performed.
For example, an organization might wish to rebuild the model using a different data mixture, change the pre-training objective, or investigate a different model architecture.
Such changes require access to upstream artifacts and training recipes that the original developers may not have released.
Thus, an organization whose development strategy relies primarily on continued training or fine-tuning successive third-party checkpoints might find itself in the model equivalent of a ``dead end'' if those releases cease or become unsuitable.
In this case, further independent progress becomes difficult without substantial effort.

As discussed above, the increasingly critical role that models play in an organization requires effective leaders to carefully weigh the benefits and costs of external dependencies, and to prepare risk mitigation strategies in advance.
In summary, fully open and sovereign models alleviate many concerns about open-weight models and allow organizations to retain control ``of their own destiny''.

\subsection{Why is this important? The Academic Perspective}
\label{section:sovereign:academic}

The previous narrative focuses on openness and sovereignty from the perspective of organizational control.
While the ability to build, deploy, and govern AI without external interference is undoubtedly important, such discussions can seem abstract from the academic perspective.\footnote{This is important because, after all, we {\it are} an academic institution.}
Building on previous discussions, we outline two concrete ways in which fully open and sovereign models are important:

\paragraph{Rigorous understanding requires transparency.}
As an academic research group, we are committed to increasing our understanding (of the world, more broadly, and LLMs, more narrowly) in a rigorous manner, and often that cannot be accomplished without transparency into the artifacts that we are studying.

In our recent NanoKnow project~\citep{Gu_etal_SIGIR2026}, we asked a simple question:\ How do LLMs know what they know?
This question is difficult to answer because pre-training data is either unknown or inaccessible, for both proprietary models and many open-weight models as well.
In a ``closed-book'' setting, we do not know if an LLM knows a fact because it simply memorized the fact from the pre-training corpus, or perhaps the model made a multi-hop inference based on other memorized facts.
In a RAG setting, where additional context is injected into the model's prompt, we cannot disentangle the effects of parametric knowledge from the grounding text provided to the model.
Similar questions apply to Hypothetical Document Embeddings (HyDE)~\citep{gao-etal-2023-precise}, a popular technique where an LLM is asked to generate a ``fake'' hypothetical document that is then used as a pivot for searching ``real'' documents in an actual collection.
When evaluating models using a benchmark test collection, we don't know if HyDE is effective because it generalized relevance across different information needs or merely because it had memorized relevant content for the test queries~\citep{yoon-etal-2025-hypothetical}.

To summarize, rigorously understanding {\it how} models answer questions is often not possible without transparency---in these cases, access to the pre-training corpus.
Transparency provides a starting point to understand the interplay between memorization and generalization in LLMs.
While transparency alone is insufficient, access to the pre-training corpus enables controlled interventions that connect data to observed behavior.
These experiments are not possible with proprietary or even open-weight models.
Unless we are able to train models from scratch, many lines of investigation will remain closed, and along with them, a complete, rigorous understanding of LLM behavior.

Transparency naturally translates into interventions that address many concerns with proprietary and open-weight LLMs.
Does a portion of the pre-training corpus violate an organization's requirements in terms of data licensing, privacy, data quality, etc.?
If so, the solution is straightforward:\ remove the data and retrain a new model.
Does the dataset used for instruction fine-tuning contain instances that are inconsistent with an organization's values?
If so, the solution is straightforward:\ remove the offending instances and retrain a new model.
Of course, it is not possible to assure that such ablations maintain model output quality, but these interventions give an organization confidence that specifically identified problems are addressed.
Such options are typically not available for proprietary and open-weight models.
Transparency with fully open and sovereign models allows one to study the impact of interventions in a controlled manner, thereby contributing to a rigorous understanding of model behavior.

\paragraph{End-to-end training provides valuable educational opportunities.}
Another cornerstone of the academic mission is to train the next generation of scholars, scientists, and researchers who will shape the future.
Richard Feynman's famous quote is apt here:\ ``What I cannot create, I do not understand''.
Our hypergraph organization of LLM training and the view of reproducibility as the traversal of hyperedges provide valuable educational opportunities for students.
There is no substitute for ``doing it oneself'', and as previously articulated elsewhere~\citep{Lin_arXiv2022}:\ reproducibility is the starting point for ``side quests'' that yield greater understanding of the phenomena under examination.
Our hypergraph formulation provides a precise way for students to think through individual steps in a much larger end-to-end training recipe and to explore how local differences may affect overall quality (e.g., in terms of retrieval metrics) via ablations and other contrastive conditions.

Starting from finished checkpoints hides many decisions that are pedagogically important, which we hope that students learn to recognize and question.
How was the pre-training corpus assembled?
How was the final pre-training data mix arrived at?
Where did the training examples come from?
What is the quality of the manually provided judgments?
What is the impact of filtering training examples?
Working through the pipeline and methodically traversing each hyperedge gives students opportunities to question decisions that are captured in existing artifacts and ponder the consequences of alternatives.
A small model trained end to end can provide educational value that is difficult to obtain from fine-tuning a much more capable model whose development is inaccessible.

The educational value also extends beyond successfully executing a recipe.
When a reproduction fails to match the reported result, students must formulate hypotheses, inspect intermediate artifacts, and design experiments that distinguish competing explanations.
They must learn to separate implementation errors from differences in configuration, data, or random variation.
With limited computing resources, they must also decide which experiments are potentially most informative.
These are important research skills, and a transparent training pipeline provides a concrete setting in which to develop them.

Finally, our hypergraph formulation supports collaboration across students and projects, and even potentially across institutions.
One student might investigate a data-processing step, while another studies a training objective, and a third explores a variant model architecture.
Shared intermediate artifacts allow students to build on each other's work without independently repeating every computation---yet retaining the ability to examine the impact of one's decision in an end-to-end task.
This provides a possible answer to the question posed in Section~\ref{section:sovereign:considerations}: Who traverses the hyperedges?
The answer:\ individual students, potentially across projects, advisors, and even institutions---in a distributed but coordinated fashion.
The resulting artifacts become both potential research contributions and educational resources for the next group of students.

\subsection{Openness {\it and} Sovereignty}

Thus far, we have advocated for fully open and sovereign models as a coherent, singular goal.
We conclude this section by distinguishing openness from sovereignty and explaining why Project Greenhouse pursues both.
The two concepts are related, but neither implies the other.
An organization can retain control over its models without making its data, code, or training recipes public.
A frontier lab can be sovereign in this sense by controlling the entire model development process from end to end while keeping its methods proprietary.
Conversely, the public availability of artifacts does not by itself ensure that another organization has the compute, expertise, or other resources needed to independently rebuild and modify a model.
Openness concerns accessibility, whereas sovereignty concerns control.

Project Greenhouse is formulated around openness {\it and} sovereignty because our goals depend simultaneously on both.
Openness ensures accessibility, which is a prerequisite for the transparency needed to facilitate explorations that lead to rigorous understanding.
As an academic institution, we are committed to sharing findings from our efforts and contributing to publicly accessible knowledge in an open manner.
On the other hand, sovereignty concerns control over dependencies, which affect the bounds of what is knowable for us.
The inability to inspect certain upstream artifacts restricts the types of explorations that we can pursue.
Access to all stages of the model development process enables the rigorous understanding of model behavior that we desire.

Together, openness {\it and} sovereignty support the rigorous understanding and educational opportunities discussed above.
Our emphasis on modest computational requirements further seeks to make these opportunities feasible for others.
However, we are cognizant of the limitations of our approach:\ it is unrealistic to imagine competing with an industry lab across a diverse range of capabilities such as general-purpose reasoning, coding, drug discovery, financial analysis, etc.
Instead, we have decided to focus on agentic search, which is broad enough to ``matter'' but yet narrowly scoped enough that we can make meaningful contributions.
We present one such small, initial contribution.

\section{Headline Results}
\label{section:headline}

\newcounter{headlinerow}
\newcounter{headlinerowgroup}
\providecommand{\theHheadlinerow}{}
\renewcommand{\theHheadlinerow}{\arabic{headlinerowgroup}.\arabic{headlinerow}}
\newcommand{\startheadlinerows}{\stepcounter{headlinerowgroup}\setcounter{headlinerow}{0}}
\newcommand{\headlinerowlabel}[1]{\refstepcounter{headlinerow}\arabic{headlinerow}\label{#1}}

This report describes how we built \ourbasereranker, our pointwise decoder-only reranker, by pre-training a causal language model from scratch on the ClimbMix corpus~\citep{diao-etal-2025-nemotron} and then fine-tuning the base checkpoint with publicly available query--document relevance annotations curated in RLHN-250K~\citep{thakur2025rlhn} using localized contrastive estimation (LCE)~\citep{gao2021lce,pradeep2022squeezing}.
This two-step recipe depends only on commonly available datasets and gives us end-to-end control over model training from initialization through relevance fine-tuning, providing a concrete step toward fully open and sovereign search models.

We were able to accomplish everything described in this report with modest computational resources, using only a handful of GPUs.
All of our supervised fine-tuning experiments can run on a single modern GPU; we typically use one or two GPUs in practice. 
For pre-training, the most resource-intensive part of our efforts, we were able to procure limited compute cycles on a single server with 8$\times$ H100 GPUs.

Our headline result involves these two model checkpoints:

\begin{itemize}[leftmargin=0.5cm]

\item \ourptmodeltt is our pre-trained causal language model checkpoint based on the nanochat codebase, detailed in Section~\ref{section:pt}.

\item \ourbasererankertt is the checkpoint of \ourbasereranker, a model soup~\citep{wortsman2022soups} formed by averaging model weights from four fine-tuning trials using the same supervised fine-tuning recipe, detailed in Section~\ref{section:sft}.

\end{itemize}

All headline results presented in this section for \ourbasereranker are from evaluating the single model soup checkpoint.

\subsection{Experimental Setup}
\label{section:headline:experimental-setup}

We implemented a relatively standard multi-stage ranking pipeline~\citep{Nogueira_etal_arXiv2019_multistageBERT} comprised of first-stage (top-$k$) retrieval that returns a ranked list of $|k|$ documents from the corpus of interest.
In our experiments, we used BM25~\citep{Robertson_Zaragoza_FnTIR2009} with $k=100$ for first-stage retrieval, as implemented with Lucene in the Pyserini toolkit~\citep{Lin_etal_SIGIR2021_Pyserini}.
Candidate documents from first-stage retrieval are then reranked, as detailed below. 

We adopt the formal definition of pointwise reranking in Section~\ref{section:agentic-search:pointwise-reranking}.
The model can be viewed as a function $\mathcal{R}(q, d_n) = s_n$ that scores each $(q, d_n)$ pair independently, where $q$ is the query and $d_n$ is a document returned by the first-stage retriever, $n \in \{ 1 \ldots k \}$.
The reranking model is trained such that reordering the list of $|k|$ documents by $s_n$ is likely to improve effectiveness as measured by some standard metric.

For evaluation, we followed standard practice and measured nDCG@10 on test collections from the TREC Deep Learning Tracks 2019--2023~\citep{Craswell_etal_TREC2019_DL_overview,Craswell_etal_TREC2020_DL_overview,Craswell_etal_TREC2021,Craswell_etal_TREC2022,Craswell_etal_TREC2023} (DL19--23 for short) and seven BEIR collections~\citep{Thakur_etal_NeurIPS2021}:\ TREC-COVID, TREC-News, Robust04, NFCorpus, SciFact, SCIDOCS, and FiQA.
We evaluated each model once per collection, and for each evaluated model, we computed an unweighted mean of its collection-level scores within each benchmark group (TREC DL and BEIR).
In our experiments, the documents and queries were truncated to 512 and 128 tokens, respectively.

We fine-tuned our reranker on RLHN-250K~\citep{thakur2025rlhn}, which includes MS MARCO, FiQA, and SCIDOCS-RR among its seven training sources.
Following the original RLHN paper, four out of our seven BEIR evaluation collections---TREC-COVID, TREC-News, Robust04, and NFCorpus---can be considered out of domain with respect to this supervised fine-tuning dataset.

We compared our \ourbasereranker with the following conditions:

\begin{itemize}[leftmargin=0.5cm]

\item \textbf{BM25:} the first-stage retrieval baseline using BM25 scores~\citep{Robertson_Zaragoza_FnTIR2009}, implemented with Pyserini~\citep{Lin_etal_SIGIR2021_Pyserini}, evaluated without reranking.

\item \textbf{Oracle reranker:} we sorted each top-100 BM25 candidate list by decreasing relevance grade from the query relevance judgments (qrels), producing an ordering in perfect agreement with the available relevance judgments.
Unjudged documents received grade zero, and documents with equal grades retained their BM25 order.
This provides the maximum nDCG@10 score achievable by reranking those candidates under the same qrels.

\item \textbf{Out-of-the-box listwise rerankers:} Gemma-4-26B-A4B~\citep{gemma-2026-gemma4} (26B total parameters, 4B active), Qwen3.8-27B~\citep{qwen38} (27B parameters), and GPT-6.1 Sol were used as listwise rerankers applied to the top-100 BM25 candidates.
For Gemma and Qwen, we used zero-shot, non-thinking generation with windows of 20 candidates, a stride of 10, and a 16,384-token context.
For GPT-6.1 Sol, we used zero-shot generation with minimal reasoning, windows of 20 candidates, and a stride of 10.
Appendix~\ref{app:listwise-prompt} provides the RankLLM~\citep{sharifymoghaddam2025rankllm} prompt used for these baselines and the Qwen-specific directive omitted for Gemma and Sol.

\item \textbf{Jev (shared-rubric scoring):} we used the \mytt{jev-latest} model from TypeSafe AI\footnote{\url{https://docs.typesafe.ai/introduction}} to score documents using the four-level TREC relevance-grade rubric, supplying the query, documents, and rubric together and sorting by each document's expected grade.
We supplied all top-100 BM25 candidates in a single request per query.
Jev appears only in the TREC DL comparisons;
Appendix~\ref{app:jev-rubric-prompt} provides the prompt and scoring rule.

\item \textbf{Jina-Reranker-v3:} a 0.6B-parameter fine-tuned listwise reranker based on Qwen3~\citep{wang2025jinarerankerv3}.\footnote{\url{https://huggingface.co/jinaai/jina-reranker-v3}}
The model jointly encodes the query and candidate documents with causal attention and computes relevance scores using contextual embeddings from each document's final token.
We supplied all top-100 BM25 candidates to the public reranking implementation, which processed them in groups of up to 64 documents.

\item \textbf{Other fine-tuned listwise rerankers:} RankVicuna~\citep{pradeep2023rankvicuna}, RankZephyr~\citep{pradeep2023rankzephyr}, and FirstMistral~\citep{chen2025accelerating}, each with 7B parameters, were applied to rerank top-100 BM25 candidates.
RankVicuna and RankZephyr were fine-tuned on MS MARCO queries and candidate passages using teacher-generated rankings:\ GPT-3.5 for RankVicuna, and GPT-3.5 followed by GPT-4 for RankZephyr.
FirstMistral ranked candidates by sorting the logits of their identifier tokens at the first output position.
These models appear only in the TREC DL comparisons, with results copied from RankLLM.

\item \textbf{MonoT5:} we used the 3B-parameter checkpoint of MonoT5~\citep{nogueira-etal-2020-document},\footnote{\url{https://huggingface.co/castorini/monot5-3b-msmarco-10k}} fine-tuned on MS MARCO for 10,000 steps.
MonoT5 scored the top-100 BM25 candidates independently with a 512-token input limit.

\item \textbf{Rank0 and Rank1:} two Qwen2.5-14B pointwise rerankers trained on the MS MARCO data mix from Rank1~\citep{weller2025rank},\footnote{\url{https://huggingface.co/datasets/jhu-clsp/rank1-training-data}} which contains query--passage pairs paired with target reasoning chains. 
Rank1 was trained to generate a reasoning chain before computing a relevance score, while Rank0~\citep{jedidi2025don} was trained on the same data but without the reasoning chains.
Both models rerank the top-100 BM25 candidates.
These models appear only in the TREC DL comparisons, with results copied from \citet{jedidi2025don}.

\item \textbf{Qwen3-Reranker:} 4B and 8B pointwise rerankers fine-tuned from Qwen3 backbones~\citep{qwen3embedding}, applied to rerank top-100 BM25 candidates.

\item \textbf{RLHN Qwen2.5-3B:} the pointwise reranker released by \citet{thakur2025rlhn},\footnote{\url{https://huggingface.co/rlhn/Qwen2.5-3B-rlhn-680K-reranker}} applied to rerank top-100 BM25 candidates.
The model was trained on the RLHN-680K data, which we discuss in Section~\ref{sec:training-set-size}.

\end{itemize}

\begin{table*}[p]
\startheadlinerows
\centering
\tableformat
\newcolumntype{R}[1]{>{\raggedleft\arraybackslash}p{#1}}
\begin{tabular}{@{}r@{\hspace{6pt}}lrR{0.85cm}R{0.85cm}R{0.85cm}R{0.85cm}R{0.85cm}R{0.85cm}@{}}
\toprule
& {\bf Model} & Size & DL19 & DL20 & DL21 & DL22 & DL23 & {\bf Mean} \\
\midrule
\headlinerowlabel{row:headline-dl:bm25} & BM25~\citep{Lin_etal_SIGIR2021_Pyserini} & --- & 0.506 & 0.480 & 0.446 & 0.269 & 0.263 & 0.393 \\
\headlinerowlabel{row:headline-dl:oracle} & Oracle reranker & --- & 0.892 & 0.871 & 0.835 & 0.678 & 0.669 & 0.789 \\
\midrule
\multicolumn{9}{@{}c@{}}{\textit{Out-of-the-box listwise rerankers}} \\
\headlinerowlabel{row:headline-dl:gemma} & Gemma-4~\citep{gemma-2026-gemma4} & 26B & 0.743 & 0.705 & 0.723 & 0.516 & 0.497 & 0.637 \\
\headlinerowlabel{row:headline-dl:qwen} & Qwen3.8~\citep{qwen38} & 27B & 0.732 & 0.705 & 0.716 & 0.520 & 0.492 & 0.633 \\
\headlinerowlabel{row:headline-dl:jev} & Jev & ? & 0.736 & 0.715 & 0.721 & 0.505 & 0.476 & 0.631 \\
\headlinerowlabel{row:headline-dl:sol} & GPT-6.1 Sol& ? & 0.750 & \textbf{0.729} & \textbf{0.725} & \textbf{0.548} & \textbf{0.513} & \textbf{0.653} \\
\midrule
\multicolumn{9}{@{}c@{}}{\textit{Fine-tuned listwise rerankers}} \\
\headlinerowlabel{row:headline-dl:rv}\rlap{$^\dagger$} & RankVicuna~\citep{pradeep2023rankvicuna} & 7B & 0.672 & 0.655 & 0.625 & 0.435 & 0.419 & 0.561 \\
\headlinerowlabel{row:headline-dl:rz}\rlap{$^\dagger$} & RankZephyr~\citep{pradeep2023rankzephyr} & 7B & 0.741 & 0.711 & 0.701 & 0.511 & 0.442 & 0.621 \\
\headlinerowlabel{row:headline-dl:mistral}\rlap{$^\dagger$} & FirstMistral~\citep{chen2025accelerating} & 7B & 0.725 & 0.701 & 0.685 & 0.489 & 0.443 & 0.609 \\
\headlinerowlabel{row:headline-dl:jina} & Jina-Reranker-v3~\citep{wang2025jinarerankerv3} & 0.6B & 0.715 & 0.696 & 0.689 & 0.503 & 0.453 & 0.611 \\
\midrule
\multicolumn{9}{@{}c@{}}{\textit{Fine-tuned pointwise rerankers}} \\
\headlinerowlabel{row:headline-dl:monot5} & MonoT5~\citep{nogueira-etal-2020-document} & 3B & 0.719 & 0.689 & 0.665 & 0.497 & 0.450 & 0.604 \\
\headlinerowlabel{row:headline-dl:rank0}\rlap{$^\dagger$} & Rank0~\citep{jedidi2025don} & 14B & 0.733 & 0.687 & 0.707 & 0.496 & 0.473 & 0.619 \\
\headlinerowlabel{row:headline-dl:rank1}\rlap{$^\dagger$} & Rank1~\citep{weller2025rank} & 14B & 0.663 & 0.650 & 0.631 & 0.448 & 0.417 & 0.562 \\
\headlinerowlabel{row:headline-dl:qwen4b} & Qwen3-Reranker~\citep{qwen3embedding} & 4B & 0.739 & 0.693 & 0.682 & 0.505 & 0.456 & 0.615 \\
\headlinerowlabel{row:headline-dl:qwen8b} & Qwen3-Reranker~\citep{qwen3embedding} & 8B & 0.731 & 0.665 & 0.663 & 0.512 & 0.454 & 0.605 \\
\headlinerowlabel{row:headline-dl:rlhn} & RLHN Qwen2.5~\citep{thakur2025rlhn} & 3B & 0.742 & 0.686 & 0.706 & 0.487 & 0.456 & 0.615 \\
\midrule
\headlinerowlabel{row:headline-dl:ours} & \ourbasereranker & 3B & \textbf{0.757} & 0.723 & 0.714 & 0.533 & 0.480 & 0.641 \\
\bottomrule
\end{tabular}
\caption{Headline results on TREC Deep Learning 2019--2023, reporting nDCG@10 on reranking top-100 BM25 candidates.
Our pointwise reranker---\ourbasereranker---is shown in row \ref{row:headline-dl:ours}.
A row number annotated with the $^\dagger$ symbol marks results reported in prior work and copied here directly; unmarked rows capture numbers from our own evaluations.
Results in rows~\ref{row:headline-dl:rv}--\ref{row:headline-dl:mistral} were taken from RankLLM~\citep{sharifymoghaddam2025rankllm}, and those in rows~\ref{row:headline-dl:rank0} and \ref{row:headline-dl:rank1} were taken from \citet{jedidi2025don}.
Mean represents the unweighted average across collections.
The highest reported score in each column, including ties, is bolded (excluding the oracle reranker).
}
\label{tab:headline-dl}
\end{table*}

\begin{table*}[p]
\startheadlinerows
\centering
\tableformat
\newcolumntype{R}[1]{>{\raggedleft\arraybackslash}p{#1}}

\begin{tabular}{@{}rlrR{0.85cm}R{0.85cm}R{0.85cm}R{0.85cm}R{0.85cm}R{0.85cm}R{0.85cm}R{0.85cm}@{}}
\toprule
 & {\bf Model} & Size & COVID & News & Robust & NFC & SciFact & SCID & FiQA & {\bf Mean} \\
\midrule
\headlinerowlabel{row:headline-beir:bm25} & BM25~\citep{Lin_etal_SIGIR2021_Pyserini} & --- & 0.595 & 0.395 & 0.407 & 0.322 & 0.679 & 0.149 & 0.236 & 0.398 \\
\headlinerowlabel{row:headline-beir:oracle} & Oracle reranker & --- & 0.975 & 0.827 & 0.837 & 0.544 & 0.927 & 0.443 & 0.590 & 0.735 \\
\midrule
\multicolumn{11}{@{}c@{}}{\textit{Out-of-the-box listwise rerankers}} \\
\headlinerowlabel{row:headline-beir:gemma} & Gemma-4~\citep{gemma-2026-gemma4} & 26B & 0.846 & 0.492 & 0.662 & 0.387 & 0.805 & 0.208 & 0.439 & 0.548 \\
\headlinerowlabel{row:headline-beir:qwen} & Qwen3.8~\citep{qwen38} & 27B & 0.849 & 0.505 & 0.647 & 0.390 & 0.803 & 0.225 & 0.462 & 0.554 \\
\headlinerowlabel{row:headline-beir:sol} & GPT-6.1 Sol& ? & 0.837 & \textbf{0.526} & \textbf{0.668} & \textbf{0.398} & \textbf{0.821} & \textbf{0.237} & \textbf{0.520} & \textbf{0.572} \\
\midrule
\multicolumn{11}{@{}c@{}}{\textit{Fine-tuned listwise rerankers}} \\
\headlinerowlabel{row:headline-beir:jina} & Jina-Reranker-v3~\citep{wang2025jinarerankerv3} & 0.6B & 0.821 & 0.459 & 0.543 & 0.359 & 0.748 & 0.215 & 0.430 & 0.511 \\
\midrule
\multicolumn{11}{@{}c@{}}{\textit{Fine-tuned pointwise rerankers}} \\
\headlinerowlabel{row:headline-beir:monot5} & MonoT5~\citep{nogueira-etal-2020-document} & 3B & 0.803 & 0.487 & 0.560 & 0.373 & 0.763 & 0.191 & 0.460 & 0.520 \\
\headlinerowlabel{row:headline-beir:qwen4b} & Qwen3-Reranker~\citep{qwen3embedding} & 4B & 0.853 & 0.497 & 0.589 & 0.374 & 0.794 & 0.235 & 0.445 & 0.541 \\
\headlinerowlabel{row:headline-beir:qwen8b} & Qwen3-Reranker~\citep{qwen3embedding} & 8B & 0.843 & 0.512 & 0.583 & 0.371 & 0.794 & 0.225 & 0.456 & 0.541 \\
\headlinerowlabel{row:headline-beir:rlhn} & RLHN Qwen2.5~\citep{thakur2025rlhn} & 3B & 0.858 & 0.465 & 0.588 & 0.312 & 0.768 & 0.217 & 0.431 & 0.520 \\
\midrule
\headlinerowlabel{row:headline-beir:ours} & \ourbasereranker & 3B & \textbf{0.859} & 0.515 & 0.592 & 0.380 & 0.790 & 0.236 & 0.455 & 0.547 \\
\bottomrule
\end{tabular}
\caption{Headline results on seven BEIR collections, reporting nDCG@10 on reranking top-100 BM25 candidates.
Our pointwise reranker---\ourbasereranker---is shown in row \ref{row:headline-beir:ours}.
All results are from our own evaluations.
Mean represents the unweighted average across collections.
The highest reported score in each column, including ties, is bolded (excluding the oracle reranker).
}
\label{tab:headline-beir}
\end{table*}

\subsection{Reranker Effectiveness and Implications}

Tables~\ref{tab:headline-dl} and~\ref{tab:headline-beir} provide our headline results, showing the effectiveness of our pointwise reranker \ourbasereranker on TREC DL and BEIR in terms of nDCG@10.
Model sizes in billions of parameters are provided as a point of reference.
In the BEIR table, COVID, News, Robust, NFC, and SCID abbreviate TREC-COVID, TREC-News, Robust04, NFCorpus, and SCIDOCS, respectively.

The comparisons include BM25, the oracle, out-of-the-box listwise models, and fine-tuned pointwise and listwise rerankers as described above.
Across both benchmark groups (TREC DL and BEIR), \ourbasereranker achieves competitive reranking effectiveness and improves over the BM25 baseline on every collection.
Furthermore, the mean nDCG@10 of \ourbasereranker exceeds those of all fine-tuned pointwise and listwise conditions that we compare against.
On TREC DL, only GPT-6.1 Sol achieves higher effectiveness.
On BEIR, \ourbasereranker achieves effectiveness comparable to Gemma-4, but appears to score lower than Qwen3.8 and GPT-6.1 Sol.
These results place our pointwise reranker, at 3B parameters, favorably among the comparison conditions, including larger models used for out-of-the-box listwise reranking.
It is not a surprise that our model is not as effective as GPT-6.1 Sol, a frontier model.

The collection-level results reveal variation that is not captured by the benchmark means:\ \ourbasereranker achieves the highest reported non-oracle nDCG@10 on DL19, even higher than GPT-6.1 Sol.
On BEIR, \ourbasereranker has higher nDCG@10 than GPT-6.1 Sol on TREC-COVID, and comparable effectiveness on SCIDOCS under the noise guideline, while Sol scores higher on the remaining collections.
Our reranker is among the strongest models on TREC-COVID, TREC-News, and SCIDOCS.

The score variability study in Section~\ref{section:sft:reproducibility} finds run-to-run sample standard deviations of approximately 0.001 for benchmark means and 0.003 on average for individual datasets.
Score differences of roughly this magnitude should be treated as experimental noise.

The oracle upper bound is shown in Row 2 of both result tables, i.e., the maximum possible scores with top-100 BM25 candidate documents.
This gap may reflect scope for reranking improvements as well as noise from incomplete or incorrect relevance judgments.
Further investigation is needed to distinguish these effects.

In support of our central claim in this report---corresponding to the first project milestone of Project Greenhouse---these results show that supervised fine-tuning using a publicly available dataset can produce a competitive reranker from a model checkpoint that we pre-train from scratch, without requiring an intermediate instruction-tuning or alignment stage, mid- or post-training, teacher supervision, synthetic data, or SFT hyperparameter tuning.
Our two-step recipe puts us in control of the entire training process, end to end.

Most importantly---and contrary to the dominant approach adopted by information retrieval researchers today---we show that it does {\it not} appear necessary to start from an existing open-weight model backbone to produce a competitive pointwise reranker.
The comparisons in Tables~\ref{tab:headline-dl} and~\ref{tab:headline-beir} assess overall reranking effectiveness across models with different sizes, data mixtures, and training and inference procedures.
Thus, we are not able to isolate any specific factors other than to provide a feasibility demonstration that provides the first step towards validating our broader thesis.

Naturally, this headline result raises many interesting questions about the effects of different backbones, different training datasets, and other architectural design choices.
We present contrastive conditions that examine some of these questions in Section~\ref{section:constrastive-conditions}, but leave many explorations for future work as we continue to pursue the broader objectives of Project Greenhouse.

\section{Pre-Training}
\label{section:pt}

Our pre-trained model backbones were produced with Andrej Karpathy's nanochat,\footnote{\url{https://github.com/karpathy/nanochat}} a compact, full-stack codebase for training decoder-only (GPT-style) language models, using the original implementation in its depth-34 configuration.
We kept nanochat's model definition, optimizer, data pipeline, and evaluation code unchanged;
modifications for pre-training our own backbones in this work involved altering the batch-size schedule.

\subsection{Experimental Setup}
\label{sec:pretraining-setup}

\paragraph{Model architecture.}
We used nanochat's depth-34 architecture without modification:\ a decoder-only transformer with 34 layers, model width 2,176, and 17 attention heads of dimension 128, trained with a 2,048-token context.

Our approach combined standard components (rotary position embeddings, QK normalization, parameter-free RMSNorm, a squared-ReLU MLP, untied input and output embeddings, and logit soft-capping) with several lightweight additions from recent speedrun-style designs.\footnote{\url{https://github.com/KellerJordan/modded-nanogpt}}
Every other layer has a value embedding, which is a per-token lookup table whose output is mixed into the attention values through a learned gate, adding token-level capacity at almost no compute cost.
Three of every four layers attend within a 512-token sliding window, while the remaining layers attend over the full context.
Learned per-layer scalars rescale the residual stream and mix the input embedding back into every layer.

In total, our model backbone has 3.29B parameters:\ 1.93B in the transformer blocks, 1.21B in the value-embedding tables, and 0.14B in the input and output embeddings.
Because value embeddings are lookups, the total per-token training compute is comparable to that of a dense model with about 2B parameters.

\paragraph{Data and tokenizer.}
We pre-trained from scratch on the ClimbMix corpus~\citep{diao-etal-2025-nemotron}, using Karpathy's document-shuffled release (\mytt{karpathy/climbmix-400b-shuffle}).
We held out the last of its 6,543 shards for validation and trained on the rest, packing documents into 2,048-token sequences with nanochat's BOS-aligned best-fit packing, in which every sequence starts at a document boundary.
Following nanochat's recipe, we trained a byte-level BPE tokenizer with a vocabulary of 32,768 entries on the first 2B characters of the training split.

\paragraph{Training.}
We kept nanochat's optimizer and learning-rate schedule unchanged:\ Muon\footnote{\url{https://kellerjordan.github.io/posts/muon/}} updates the weight matrices of the transformer blocks and AdamW updates the embeddings and scalar parameters.
Our only modifications were to change the learning rate and the batch size schedule.

Unlike nanochat, which scales its peak learning rates with the square root of the batch size, we kept their peak values unchanged as the batch grows, so each increase cuts the gradient noise much as a learning rate decay would.
Early in pre-training, since the model is not very good, even a small data batch clearly shows the direction of an effective update; 
in these situations, a larger batch may provide limited additional benefit per token, while consuming more tokens per step.
Thus, using small data batches at the beginning of pre-training is more data efficient.

\begin{figure}[t]
\centering
\includegraphics[width=0.5\linewidth]{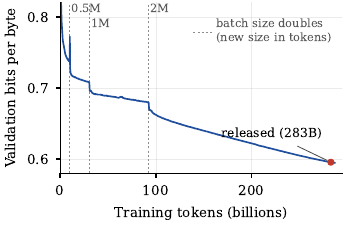}
\caption{Validation loss on held-out ClimbMix during pre-training.}
\label{fig:loss}
\end{figure}

As the loss decreases, however, the signal-to-noise ratio of each batch decreases, so larger batches start to become worthwhile~\citep{mccandlish-etal-2018-empirical}.
Therefore, we doubled the batch size three times, from 262,144 to 524,288, 1,048,576, and 2,097,152 tokens over the course of training, which also reduced the total number of optimizer steps.
We kept the learning rates fixed throughout the pre-training process, so each increase of the batch size cuts the gradient noise, much as a learning-rate decay would~\citep{smith-etal-2018-dont}.
These batch size increases correspond to the discontinuities in the validation loss plot in Figure~\ref{fig:loss}.
For example, we see that the validation loss drops suddenly, with a brief spike at the first batch size increase.

In total, we pre-trained for exactly one pass over the entire corpus (comprising 287B tokens), on a single server with 8$\times$ NVIDIA H100 GPUs with FP8 matrix multiplications.
The entire process consumed roughly 1,600 GPU-hours in this configuration.

\paragraph{Bidirectional variant.}
Our GPU budget did not allow for a second pre-training run from scratch, so we adapted the causal model to bidirectional attention following LLM2Vec~\citep{behnamghader-etal-2024-llm2vec}.
Starting from the causal checkpoint at 245B tokens, we continued pre-training on the remaining 42B tokens with masked next-token prediction (MNTP). 
MNTP predicted each masked token from the preceding position while allowing attention to context on both sides.
Each prediction therefore used the same position as in causal next-token training.
Since the checkpoint had no saved optimizer state, we raised the learning rate from zero to 0.26 times the peak over the first 200 steps, where the causal run's schedule had reached, and then followed that schedule's linear decay to 0.05 times the peak.
Adaptation used approximately 19,900 steps, a 20\% masking rate, and symmetric attention windows.
The adaptation consumed approximately 340 H100-hours.

\subsection{Model Evaluation}

It is, of course, desirable to assess the quality of a pre-trained model intrinsically, without consideration of any specific downstream task. 
To evaluate base pre-trained models without post-training, it is necessary to use benchmarks that do not require instruction following:

\begin{itemize}[leftmargin=0.5cm]

\item \textbf{CORE}~\citep{dclm}: 22 in-context learning tasks from DataComp-LM, covering world knowledge, commonsense reasoning, language understanding, reading comprehension, and symbolic problem solving. 
CORE averages accuracy on each task after rescaling the score so that random guessing scores 0 and perfect accuracy scores 1.
We used nanochat's implementation with up to 500 test questions per task.

\item \textbf{MMLU}~\citep{hendrycks-etal-2021-measuring}: four-option multiple-choice questions on 57 subjects.
We used 5-shot prompts and picked the answer letter (A--D) that the model rates as most likely, using the LM Evaluation Harness.\footnote{\url{https://github.com/eleutherai/lm-evaluation-harness}}

\item \textbf{MMLU-Pro}~\citep{wang-etal-2024-mmlu}: a harder, ten-option version of MMLU, scored the same way.
We do not use the more common chain-of-thought setting, where the model writes out its reasoning before answering, because its prompts do not fit in our 2,048-token context.

\end{itemize}
\noindent Models evaluated locally used the same evaluation code, with each tokenizer's default special tokens and each model's own context length.
Our bidirectional variant, which would otherwise see the answer, is scored with the answer tokens masked, as during model pre-training.
In cases where a model developer directly reports the relevant metric, we use the published scores;
otherwise, we performed evaluations ourselves.

To confirm the correctness of our evaluation code, we validated against published results.
On seven reference models with published CORE scores, our CORE is within 0.011 of published values.
These are:\ GPT-2 at four sizes~\citep{radford-etal-2019-language}, OLMo-1B~\citep{groeneveld-etal-2024-olmo}, phi-1.5~\citep{li-etal-2023-textbooks}, and Qwen2-1.5B~\citep{yang-etal-2024-qwen2}.
Our MMLU and MMLU-Pro scores are within 0.003 of the stock harness, which in turn reproduces published chain-of-thought MMLU-Pro scores to within 0.02.
CORE fluctuates slightly late in training, so we released the causal checkpoint with the highest CORE (283B tokens) rather than the final model (287B tokens, 0.004 lower).
It is this model checkpoint that we release and evaluate throughout this report.

Table~\ref{tab:pretrain-compare} compares our models with well-known base models of up to about 7B parameters released in the past year, grouped by whether their pre-training corpus is public.
For reference, the table also includes AI2's two latest OLMo base models, whose pre-training corpus is publicly available, but which are only available at 7B.

Based on our evaluation, our causal model scores much higher than nanochat-d34, which was trained with an earlier version of the same codebase on 89B tokens of FineWeb-Edu.
This contrast shows the importance of the pre-training corpus and token budget.
Other interesting observations:\ 
Puro 2B achieves a CORE score similar to ours with about five times as many training tokens.
Apart from nanochat-d34, Puro 2B, and OPEN-1B, every model, including all models with undisclosed data, scores higher than ours, but each was trained on 1.7T to 34T tokens or does not report its budget.
The models shown in Table~\ref{tab:pretrain-compare} differ in data, architecture, and training recipe, so the comparison places our model among existing checkpoints rather than isolating any single factor.
Our primary goal here is to support the claim that our model checkpoint, pre-trained from scratch ourselves, provides a ``reasonable'' starting point for subsequent fine-tuning to yield a competitive pointwise reranker.

\begin{table*}[t]
\centering
\tableformat
\begin{tabular}{@{}rllrrrrr@{}}
\toprule
& Model & Data & Params & Tokens & CORE & MMLU & MMLU-Pro \\
\midrule
1 & \textbf{gaggle-nanochat-pretrained} & ClimbMix & 3.3B & 283B & 0.458 & 0.493 & 0.178 \\
2 & \textbf{gaggle-nanochat-pretrained-bidirectional} \mbox{ } \rlap{$^\ddagger$} & ClimbMix & 3.3B & 287B & 0.339 & 0.453 & 0.187 \\
\midrule
\multicolumn{8}{@{}c@{}}{\textit{Open pre-training data}} \\
3 & nanochat-d34 & FineWeb-Edu       & 2.2B & 89B   & 0.338\rlap{$^\dagger$} & 0.257 & 0.115 \\
4 & OPEN-1B~\citep{open-1b} & OPEN-1B mix       & 1.6B & 400B  & 0.239 & 0.259\rlap{$^\dagger$} & 0.117\rlap{$^\dagger$} \\
5 & Puro 2B~\citep{luo-etal-2026-puro} & Puro mix          & 2.0B & 1.4T  & 0.464 & 0.552 & 0.235 \\
6 & Apertus v1.1 4B~\citep{panferov-etal-2026-apertus} & Apertus corpus    & 3.8B & 1.7T  & 0.495 & 0.573 & 0.231 \\
7 & K2-Horizon 3.7B & TxT360-v2 mix     & 5.1B & 22.9T & 0.554 & 0.664\rlap{$^\dagger$} & 0.380 \\
8 & MiniCPM5-2B & Ultra-FineWeb mix & 2.5B & --    & 0.500 & 0.653 & 0.384 \\
\midrule
\multicolumn{8}{@{}c@{}}{\textit{Open pre-training data, 7B (latest OLMo, for reference)}} \\
9 & Olmo 3 7B~\citep{olmo3} & Dolma 3           & 7.3B & 6T  & 0.562 & 0.669\rlap{$^\dagger$} & 0.373\rlap{$^\dagger$} \\
10 & Olmo Hybrid 7B~\citep{merrill-etal-2026-olmo} & Dolma 3           & 7.4B & 6T    & 0.579 & 0.685 & 0.417\rlap{$^\dagger$} \\
\midrule
\multicolumn{8}{@{}c@{}}{\textit{Undisclosed pre-training data}} \\
11 & Qwen3.5-4B~\citep{qwen-2026-qwen35} & Undisclosed       & 4.2B & --    & 0.589 & 0.731 & 0.487 \\
12 & Qwen3.5-2B~\citep{qwen-2026-qwen35} & Undisclosed       & 1.9B & --    & 0.484 & 0.540 & 0.367 \\
13 & Gemma-4-E2B~\citep{gemma-2026-gemma4} & Undisclosed       & 4.6B & --    & 0.507 & 0.577 & 0.247 \\
14 & Granite 4.1 3B & Undisclosed       & 3.4B & 15T   & 0.555 & 0.665\rlap{$^\dagger$} & 0.336 \\
15 & LFM2.5-2.6B & Undisclosed       & 2.7B & 34T   & 0.485 & 0.637 & 0.371 \\
\bottomrule
\end{tabular}
\caption{Overview of a selection of recent pre-trained base models.
Rows 1 and 2 are our causal model trained from scratch and its bidirectional variant, adapted from the causal checkpoint with masked next-token prediction.
For all scores, higher is better.
$^\dagger$ indicates scores reported by the model developer and copied here directly; other scores are from our own evaluations.
$^\ddagger$ indicates scoring with masked prediction.
Params include embedding tables.
}
\label{tab:pretrain-compare}
\end{table*}

\subsection{Released Artifacts}

The pre-training phase of our project concluded with two model checkpoints that we share and build on in the supervised fine-tuning step:

\begin{itemize}[leftmargin=0.5cm]

\item \ourptmodeltt is our causal pre-trained checkpoint, as detailed in this section.

\item \ourbidirptmodeltt{} is the bidirectional variant described above (starting with causal modeling and switching to bidirectional modeling).
We also release the modified attention code needed to perform inference using this variant.

\end{itemize}

\section{Supervised Fine-Tuning}
\label{section:sft}

This section details the supervised fine-tuning recipe applied to our pre-trained checkpoint to produce a competitive pointwise reranker.
We provide an exploration of model score variability across four training runs and describe the final model soup \ourbasereranker used for the headline results in Section~\ref{section:headline}.

% Number data rows within each SFT table and provide stable cross-references.
\newcounter{sftrow}
\newcounter{sftrowgroup}
\providecommand{\theHsftrow}{}
\renewcommand{\theHsftrow}{\arabic{sftrowgroup}.\arabic{sftrow}}
\newcommand{\startsftrows}{\stepcounter{sftrowgroup}\setcounter{sftrow}{0}}
\newcommand{\sftrowlabel}[1]{\refstepcounter{sftrow}\arabic{sftrow}\label{#1}}
\newcommand{\sftresultref}[2]{Table~\ref{#1}, row~\ref{#2}}

\subsection{Experimental Setup}
\label{section:sft_config}

Our reranker adopts the formal definition of pointwise reranking from Section~\ref{section:agentic-search:pointwise-reranking} and the setup described in our headline results from Section~\ref{section:headline:experimental-setup}.
We do not repeat the formalism here.

\begin{figure}[!tp]
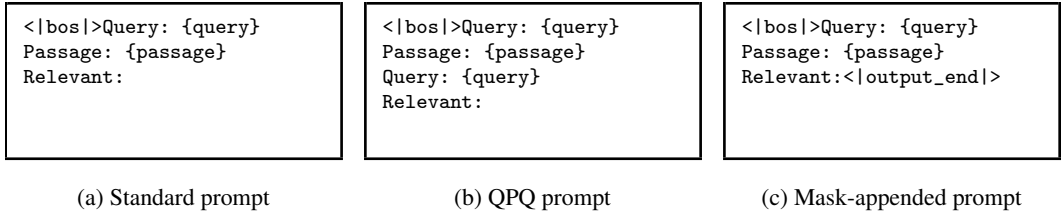

\centering
\begin{subfigure}[t]{0.32\textwidth}
\begin{mdframed}[font=\footnotesize, roundcorner=6pt, linewidth=1pt, innerleftmargin=6pt, innerrightmargin=6pt, innertopmargin=5pt, innerbottommargin=5pt]
\begin{minipage}[t][5\baselineskip][t]{\linewidth}
\begin{verbatim}
<|bos|>Query: {query}
Passage: {passage}
Relevant:
\end{verbatim}
\end{minipage}
\end{mdframed}
\caption{Standard prompt}
\label{fig:reranker-prompt-standard}
\end{subfigure}
\hfill
\begin{subfigure}[t]{0.32\textwidth}
\begin{mdframed}[font=\footnotesize, roundcorner=6pt, linewidth=1pt, innerleftmargin=6pt, innerrightmargin=6pt, innertopmargin=5pt, innerbottommargin=5pt]
\begin{minipage}[t][5\baselineskip][t]{\linewidth}
\begin{verbatim}
<|bos|>Query: {query}
Passage: {passage}
Query: {query}
Relevant:
\end{verbatim}
\end{minipage}
\end{mdframed}
\caption{QPQ prompt}
\label{fig:reranker-prompt-qpq}
\end{subfigure}
\hfill
\begin{subfigure}[t]{0.32\textwidth}
\begin{mdframed}[font=\footnotesize, roundcorner=6pt, linewidth=1pt, innerleftmargin=6pt, innerrightmargin=6pt, innertopmargin=5pt, innerbottommargin=5pt]
\begin{minipage}[t][5\baselineskip][t]{\linewidth}
\begin{verbatim}
<|bos|>Query: {query}
Passage: {passage}
Relevant:<|output_end|>
\end{verbatim}
\end{minipage}
\end{mdframed}
\caption{Mask-appended prompt}
\label{fig:reranker-prompt-mask}
\end{subfigure}
\caption{Prompts used for both reranker supervised fine-tuning and evaluation.
(a) The standard prompt is used for \ourbasereranker and all other contrastive studies by default.
(b) The QPQ prompt is used for the causal query-repetition condition; both occurrences contain the same truncated query.
(c) The mask-appended prompt is used for the Gaggle (MNTP) mask condition; \mytt{<|output\_end|>} is the MNTP mask token.
With the single-position readout, relevance is scored using the \mytt{\textvisiblespace{}true} minus \mytt{\textvisiblespace{}false} logits at the colon after \mytt{Relevant:}, immediately before the mask in (c).}
\label{fig:reranker-prompt}
\end{figure}

\paragraph{Model and scoring.}
The \ourptmodeltt checkpoint, our 3.29B-parameter causal language model pre-trained on ClimbMix (see Section~\ref{section:pt}), provides a starting point for \ourbasereranker.
The reranker scores each query--passage pair independently, $\mathcal{R}(q, d_n) = s_n$, using the ``standard'' prompt in Figure~\ref{fig:reranker-prompt}, panel (a).
Our training prompts used the label \mytt{Passage:} for the candidate text.
However, we use ``document'' and ``passage'' interchangeably throughout the paper.
The relevance score $s$ is computed as the difference between the next-token logits for \mytt{\textvisiblespace{}true} and \mytt{\textvisiblespace{}false}, each a single token that includes the leading space shown as \mytt{\textvisiblespace}:
\begin{equation}
\label{eq:sft-score}
s(q,d)=\ell_{\mathrm{true}}(q,d)-\ell_{\mathrm{false}}(q,d).
\end{equation}
We retained the pre-trained language-model output rows for these two tokens (2$\times$2176$~=~$4,352 parameters) and fine-tuned them jointly with the backbone.
The remaining output rows were discarded, leaving approximately 3.22B parameters in the reranker, compared with 3.29B parameters in the original language model.
No classification head was added, and all retained parameters were initialized from the pre-trained checkpoint.
Queries and passages were truncated to 128 and 512 tokens, respectively, during both training and evaluation.

\paragraph{Training data.}
It is well known~\citep{Nogueira:1901.04085:2019} that a high-quality pointwise reranker can be fine-tuned from a pre-trained transformer backbone using a collection of query--document relevance pairs.
We adopt this approach, but the biggest variable remains the data recipe:\ exactly what training data and what data mix.

In this work, we used RLHN-250K, a curated subset of seven retrieval datasets from the BGE training collection~\citep{thakur2025rlhn}.
Its construction uses a small LLM to identify potential false negatives and a stronger LLM to verify them; confirmed false negatives are relabeled as positives, and instances with an excessive number of false negatives are discarded.
The baseline dataset used in training \ourbasereranker therefore already used filtered and relabeled data; further experiments in Section~\ref{sec:self-filtering-distillation} examined the effects of additional curation.

\paragraph{Training objective.}
We used localized contrastive estimation (LCE), introduced by \citet{gao2021lce} and subsequently studied by \citet{pradeep2022squeezing}.
Both studies found LCE more effective than pointwise cross-entropy (CE), which treats each query--document pair independently.
LCE applies a softmax over one positive passage and $K$ sampled hard negatives for the same query:
\begin{equation}
\label{eq:sft-lce}
\mathcal{L}_{\mathrm{LCE}}
=-\log\frac{\exp(s(q,d^+)/\tau)}
{\exp(s(q,d^+)/\tau)+\sum_{i=1}^{K}\exp(s(q,d_i^-)/\tau)}.
\end{equation}
Our comparison with pointwise cross-entropy in Section~\ref{sec:ce-loss-ablation} further supports this choice under the tested configuration.
We fixed the softmax temperature at $\tau=1$ in all LCE training runs.

\paragraph{Configuration.}
We fine-tuned the reranker for one epoch using AdamW~\citep{loshchilov2019adamw} with 150 steps of linear warm-up to a peak learning rate of $5\times10^{-5}$, followed by cosine decay.
We used a common optimization recipe for the contrastive studies, changing the factors specified by each experiment.
We did not perform hyperparameter tuning for the baseline or the contrastive studies in Section~\ref{section:constrastive-conditions}, except for the diagnostic comparison of peak learning rates for Gemma and MiniCPM in Section~\ref{section:different-backbones}.
Appendix~\ref{app:sft-configuration} records data sampling and microbatching, detailed optimization and implementation settings, and the shared and per-run configurations.

\subsection{Run-to-Run Variability and Model Averaging}
\label{section:sft:reproducibility}

We performed four baseline supervised fine-tuning runs to characterize run-to-run variation in ranking effectiveness. 
Each run started from the same causal pre-trained checkpoint and used the RLHN-250K data, following the same recipe detailed here.

To provide more details:\
runs 1, 3, and 4 used seed 17, while run 2 used seed 18.
Runs 1--3 used two RTX PRO 6000 Blackwell GPUs; run 4 used two H100 GPUs with gradient caching.
The runs shared the optimization recipe detailed here but differed in execution environment and microbatch configuration, as explained in Appendix~\ref{app:sft-configuration}.
Our comparison therefore measures variation across multiple trials that include changes in seed and execution settings.

\begin{table*}[t]
\startsftrows
\centering
\tableformat
\newcolumntype{R}[1]{>{\raggedleft\arraybackslash}p{#1}}
\begin{subtable}[t]{\textwidth}
\centering
\tableformat
\captionsetup{position=bottom,font={footnotesize,it},justification=centering,singlelinecheck=false,skip=4pt}

\begin{tabular}{@{}rlrR{0.85cm}R{0.85cm}R{0.85cm}R{0.85cm}R{0.85cm}R{0.85cm}@{}}
\toprule
& Model & Seed & DL19 & DL20 & DL21 & DL22 & DL23 & {\bf Mean} \\
\midrule
\multicolumn{9}{@{}c@{}}{\textit{Individual runs}} \\
\sftrowlabel{row:repro:baseline} & Baseline -- run 1 & 17 & 0.755 & 0.720 & \textbf{0.715} & \textbf{0.533} & \textbf{0.485} & \textbf{0.641} \\
\sftrowlabel{row:repro:seed18} & Baseline -- run 2 & 18 & \textbf{0.758} & \textbf{0.723} & 0.707 & 0.531 & 0.480 & 0.640 \\
\sftrowlabel{row:repro:nour} & Baseline -- run 3 & 17 & 0.755 & 0.720 & 0.714 & \textbf{0.533} & 0.481 & \textbf{0.641} \\
\sftrowlabel{row:repro:lingwei} & Baseline -- run 4 & 17 & 0.750 & 0.720 & \textbf{0.715} & 0.528 & 0.480 & 0.638 \\
\midrule
\multicolumn{9}{@{}c@{}}{\textit{Soup checkpoint}} \\
\sftrowlabel{row:repro:average} & \ourbasereranker & --- & 0.757 & \textbf{0.723} & 0.714 & \textbf{0.533} & 0.480 & \textbf{0.641} \\
\midrule
\multicolumn{9}{@{}c@{}}{\textit{Stats summary}} \\
\multicolumn{3}{l}{Mean of runs 1--4} & 0.754 & 0.721 & 0.712 & 0.531 & 0.481 & 0.640 \\
\multicolumn{3}{l}{Sample SD of runs 1--4} & 0.003 & 0.001 & 0.004 & 0.002 & 0.002 & 0.001 \\
\bottomrule
\end{tabular}
\caption{TREC Deep Learning}
\label{tab:sft-reproducibility-dl}
\end{subtable}
\medskip

\begin{subtable}[t]{\textwidth}
\centering
\tableformat
\captionsetup{position=bottom,font={footnotesize,it},justification=centering,singlelinecheck=false,skip=4pt}

\begin{tabular}{@{}rlrR{0.85cm}R{0.85cm}R{0.85cm}R{0.85cm}R{0.85cm}R{0.85cm}R{0.85cm}R{0.85cm}@{}}
\toprule
& Model & Seed & COVID & News & Robust & NFC & SciFact & SCID & FiQA & {\bf Mean} \\
\midrule
\multicolumn{11}{@{}c@{}}{\textit{Individual runs}} \\
\ref{row:repro:baseline} & Baseline -- run 1 & 17 & 0.858 & 0.510 & 0.591 & 0.381 & 0.784 & 0.233 & 0.451 & 0.544 \\
\ref{row:repro:seed18} & Baseline -- run 2 & 18 & 0.852 & 0.512 & 0.584 & \textbf{0.383} & 0.781 & 0.235 & 0.448 & 0.542 \\
\ref{row:repro:nour} & Baseline -- run 3 & 17 & \textbf{0.862} & \textbf{0.516} & 0.591 & 0.380 & 0.784 & 0.234 & 0.452 & 0.545 \\
\ref{row:repro:lingwei} & Baseline -- run 4 & 17 & 0.860 & 0.515 & \textbf{0.592} & 0.377 & 0.783 & 0.234 & 0.454 & 0.545 \\
\midrule
\multicolumn{11}{@{}c@{}}{\textit{Soup checkpoint}} \\
\ref{row:repro:average} & \ourbasereranker & --- & 0.859 & 0.515 & \textbf{0.592} & 0.380 & \textbf{0.790} & \textbf{0.236} & \textbf{0.455} & \textbf{0.547} \\
\midrule
\multicolumn{11}{@{}c@{}}{\textit{Stats summary}} \\
\multicolumn{3}{l}{Mean of runs 1--4} & 0.858 & 0.513 & 0.589 & 0.380 & 0.783 & 0.234 & 0.451 & 0.544 \\
\multicolumn{3}{l}{Sample SD of runs 1--4} & 0.004 & 0.003 & 0.003 & 0.002 & 0.001 & 0.001 & 0.002 & 0.001 \\
\bottomrule
\end{tabular}
\caption{BEIR}
\label{tab:sft-reproducibility-beir}
\end{subtable}
\caption{Analysis of run-to-run variability and model averaging (nDCG@10).
All models have 3.22B parameters; the four individual runs used the same training recipe, with differences in random seed, GPU hardware, microbatch token budget, and resulting update count.
All models rerank top-100 BM25 candidates.
Rows 1--4 represent individual trials; row 5 is the soup checkpoint used in our headline results.
Summary rows report the mean and sample standard deviation (SD) across rows 1--4 only.
The highest score in each column is marked in bold (including ties, excluding summary rows).}
\label{tab:sft-reproducibility}
\end{table*}

Finally, we formed a model soup~\citep{wortsman2022soups} by averaging all parameters across the final model checkpoints from the four individual runs above, with a weight of 0.25 per checkpoint.
The model soup corresponds to the checkpoint \ourbasererankertt, which is used to produce our headline result in Section~\ref{section:headline}.

Our model soup and individual models are evaluated in exactly the same way with results shown in Table~\ref{tab:sft-reproducibility}.
To be explicit:\ the soup scores were obtained from this single averaged model (i.e., this is {\it not} so-called results averaging).
The variability of individual runs is summarized in terms of the sample standard deviation across the four individual models.

The four individual runs achieve similar ranking effectiveness:
means of 0.640$\pm$0.001 on TREC DL and 0.544$\pm$0.001 on BEIR, where $\pm$ denotes the sample standard deviation.
We also computed the sample standard deviation separately for each test collection; these values range from 0.001 to 0.004, with an average of 0.003 across the 12 collections.
We use these standard deviation measurements as our guideline for run-to-run noise when interpreting differences in the contrastive studies.
That is, we consider differences that are roughly below this threshold to be experimental noise.

The soup model achieved a mean nDCG@10 of 0.641 on TREC DL and 0.547 on BEIR, exceeding the corresponding averages of the individual runs by 0.001 and 0.003.
Its TREC DL mean matched the strongest individual run, while its BEIR mean exceeded all four individual runs.
Although these gains are close to the observed run-to-run noise scale, the results suggest a potential benefit from parameter averaging, even when all model instances follow the same training recipe.
This appears consistent with previous studies exploring model soups~\citep{wortsman2022soups}.

\section{Contrastive Conditions}
\label{section:constrastive-conditions}

In this section, we examine how modeling and training choices affect reranking effectiveness.
Our experiments considered a number of different factors against baseline run 1 of our \ourbasereranker, as described in Section~\ref{section:sft:reproducibility}.
To be clear, we do {\it not} compare against the model soup, which would not be methodologically fair.

We interpret score differences descriptively using the per-trial variability measured in Section~\ref{section:sft:reproducibility} as a reference. 
Specifically, we treat changes on the order of 0.001 in benchmark means or 0.003 on individual collections as experimental noise.
Comparisons use the unrounded evaluation scores.

\subsection{The Effects of Different Backbones}
\label{section:different-backbones}

\subsubsection{Experimental Setup}

Our headline result demonstrates that we are able to fine-tune a pointwise reranker (Section~\ref{section:sft}) from our own pre-trained model checkpoint (Section~\ref{section:pt}).
A natural follow-up question is:\ What about other model backbones?
How much of our success is attributable to the choice of model backbones?

To answer this question, we applied exactly the same supervised fine-tuning recipe (as described in Section~\ref{section:sft}) but starting from different open-weight backbones:\ nanochat-d34,\footnote{\url{https://huggingface.co/karpathy/nanochat-d34}} Gemma-4-E2B,\footnote{\url{https://huggingface.co/google/gemma-4-E2B}} Qwen3.5-4B-Base,\footnote{\url{https://huggingface.co/Qwen/Qwen3.5-4B-Base}} and MiniCPM5-2B-Base.\footnote{\url{https://huggingface.co/openbmb/MiniCPM5-2B-Base}}

All training runs used causal attention, the standard prompt, the LCE loss function, and the last non-padded token for scoring.
We fine-tuned each model's text parameters on RLHN-250K, matching the optimization hyperparameters and input-length limits used for our headline results.

Additional details for different backbones:\
For MiniCPM, we used \mytt{openbmb/MiniCPM5-2B-Base}, the model checkpoint with pre-training only, rather than the mid-trained release.
For the multimodal backbones Gemma-4-E2B and Qwen3.5-4B-Base, we discarded the vision encoders and Gemma's audio encoder, leaving approximately 4.6B and 4.2B trainable parameters, respectively; Gemma's count includes its per-layer embedding tables.
Reported model parameter sizes in our results table exclude untied output layers, since scoring uses only the two answer-token rows; for example, nanochat-d34 has 2.1B parameters under this convention and 2.2B if we include its output layer.
Each backbone retained its native architecture and tokenizer.

We first fine-tuned every backbone with the peak learning rate used in our headline results, which is $5{\times}10^{-5}$.
Additional analyses showed that this peak was too high for Gemma-4-E2B and MiniCPM, so we reran both with a peak learning rate of $1{\times}10^{-5}$, keeping the warm-up and decay schedule unchanged (details below).

\subsubsection{Results and Discussion}

\begin{table*}[t]
\startsftrows
\centering
\tableformat
\newcolumntype{R}[1]{>{\raggedleft\arraybackslash}p{#1}}
\begin{tabular}{@{}rlllR{1.25cm}R{1.25cm}@{}}
\toprule
& Backbone & Size & LR & TREC DL & BEIR \\
\midrule
\sftrowlabel{row:backbone:ours} & gaggle-nanochat-pretrained \mbox{ \qquad } & 3.2B & $5{\times}10^{-5}$ & 0.641 & 0.544 \\
\sftrowlabel{row:backbone:nanochat} & nanochat-d34 & 2.1B & $5{\times}10^{-5}$ & 0.621 & 0.521 \\
\sftrowlabel{row:backbone:gemma} & Gemma-4-E2B & 4.6B & $5{\times}10^{-5}$ & 0.627 & 0.537 \\
\sftrowlabel{row:backbone:qwen} & Qwen3.5-4B-Base & 4.2B & $5{\times}10^{-5}$ & 0.626 & 0.530 \\
\sftrowlabel{row:backbone:minicpm} & MiniCPM5-2B-Base & 2.2B & $5{\times}10^{-5}$ & 0.627 & 0.530 \\
\midrule
\sftrowlabel{row:backbone:gemma-lowlr} & Gemma-4-E2B & 4.6B & $1{\times}10^{-5}$ & 0.640 & 0.546 \\
\sftrowlabel{row:backbone:minicpm-lowlr} & MiniCPM5-2B-Base & 2.2B & $1{\times}10^{-5}$ & {\bf 0.642} & {\bf 0.551} \\
\bottomrule
\end{tabular}
\caption{Effects of different backbones (mean nDCG@10).
All configurations are trained with the same supervised fine-tuning recipe.
LR denotes the peak learning rate; other hyperparameters follow the headline recipe, and hardware and microbatch differences are documented in Appendix~\ref{app:sft-configuration}.
All configurations rerank top-100 BM25 candidates.
Row~1 is baseline run~1 of \ourbasereranker.
The highest score in each column is marked in bold.}
\label{tab:backbone-results}
\end{table*}

Experimental results of reranker effectiveness with different backbones are shown in Table~\ref{tab:backbone-results}.
Using the shared supervised fine-tuning recipe, baseline run~1 of \ourbasereranker achieves a higher mean nDCG@10 than the other backbones on both TREC DL and BEIR, with differences exceeding the run-to-run noise guideline.
Outperforming nanochat-d34 did not come as a surprise, since our backbone was pre-trained from a better data mix, but results for the other backbones were surprising.
In particular, given that Gemma and MiniCPM achieve higher CORE, MMLU, and MMLU-Pro scores (see Table~\ref{tab:pretrain-compare}), we found it perplexing that using these backbones as a starting point for supervised fine-tuning resulted in a less effective reranker.

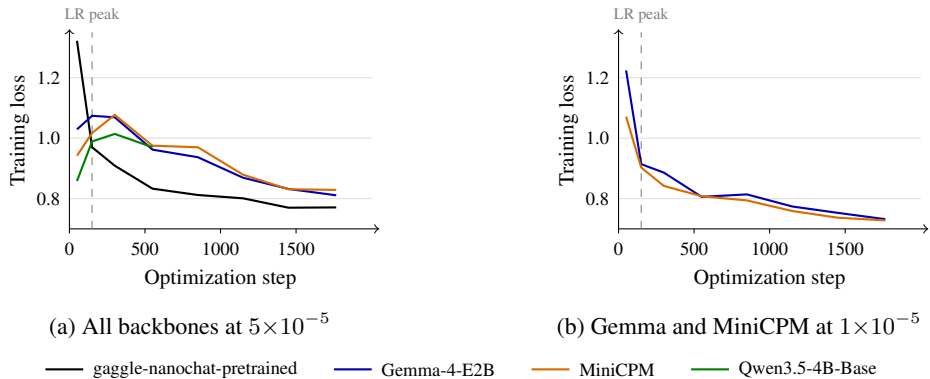
\begin{figure}[t]
\centering
\tikzset{
  ours/.style={black, thick},
  gem/.style={blue!70!black, thick},
  mini/.style={orange!85!black, thick},
  qwen/.style={green!50!black, thick},
  lossaxes/.style={x=0.002cm, y=4cm, font=\scriptsize},
}
\newcommand{\lossframe}{%
  \foreach \yv in {0.8,1.0,1.2}
    {\draw[gray!25] (0,\yv) -- (2000,\yv); \node[left] at (0,\yv) {\yv};}
  \foreach \xv in {0,500,1000,1500}
    {\draw (\xv,0.70) -- (\xv,0.685); \node[below] at (\xv,0.685) {\xv};}
  \draw[->] (0,0.70) -- (2050,0.70);
  \draw[->] (0,0.70) -- (0,1.36);
  \draw[gray, dashed] (150,0.70) -- (150,1.35) node[above, font=\tiny] {LR peak};
  \node[below=10pt] at (1000,0.685) {\footnotesize Optimization step};
  \node[rotate=90] at (-330,1.025) {\footnotesize Training loss};}
\begin{subfigure}[t]{0.48\textwidth}
\centering
\begin{tikzpicture}[lossaxes]
\begin{scope}[shift={(0,-0.70)}]
  \lossframe
  \draw[ours] plot coordinates {(50,1.322) (150,0.970) (300,0.909) (550,0.833) (850,0.812) (1150,0.801) (1450,0.770) (1766,0.771)};
  \draw[gem]  plot coordinates {(50,1.029) (150,1.074) (300,1.069) (550,0.962) (850,0.937) (1150,0.869) (1450,0.832) (1766,0.811)};
  \draw[mini] plot coordinates {(50,0.942) (150,1.017) (300,1.077) (550,0.975) (850,0.970) (1150,0.879) (1450,0.831) (1766,0.829)};
  \draw[qwen] plot coordinates {(50,0.858) (150,0.989) (300,1.014) (550,0.970)};
\end{scope}
\end{tikzpicture}
\caption{All backbones at $5{\times}10^{-5}$}
\label{fig:lr-backbones-loss}
\end{subfigure}
\hfill
\begin{subfigure}[t]{0.48\textwidth}
\centering
\begin{tikzpicture}[lossaxes]
\begin{scope}[shift={(0,-0.70)}]
  \lossframe
  \draw[gem]  plot coordinates {(50,1.224) (150,0.914) (300,0.886) (550,0.806) (850,0.814) (1150,0.774) (1450,0.753) (1766,0.732)};
  \draw[mini] plot coordinates {(50,1.071) (150,0.903) (300,0.842) (550,0.808) (850,0.794) (1150,0.759) (1450,0.737) (1766,0.728)};
\end{scope}
\end{tikzpicture}
\caption{Gemma and MiniCPM at $1{\times}10^{-5}$}
\label{fig:lr-backbones-lowlr}
\end{subfigure}

\smallskip
\begin{tikzpicture}[font=\scriptsize]
  \draw[ours] (0,0) -- (0.5,0) node[right, text=black] (l1) {gaggle-nanochat-pretrained};
  \draw[gem]  ([xshift=0.35cm]l1.east) -- ++(0.5,0) node[right, text=black] (l2) {Gemma-4-E2B};
  \draw[mini] ([xshift=0.35cm]l2.east) -- ++(0.5,0) node[right, text=black] (l3) {MiniCPM};
  \draw[qwen] ([xshift=0.35cm]l3.east) -- ++(0.5,0) node[right, text=black] {Qwen3.5-4B-Base};
\end{tikzpicture}
\caption{Training loss on identical batches at peak learning rates $5{\times}10^{-5}$ (a) and $1{\times}10^{-5}$ (b); bins widen over training to smooth per-batch noise.}
\label{fig:lr-backbones}
\end{figure}

We followed up and examined the training dynamics of Gemma and MiniCPM to better understand their lower reranking effectiveness with the shared recipe.
All backbones initially used the same peak learning rate of $5\times10^{-5}$, which we had used for our headline results without hyperparameter tuning.
On identical batches, training losses for Gemma and MiniCPM increased as the learning rate reached its peak, while losses from our model continued to decrease, as shown in Figure~\ref{fig:lr-backbones-loss}.

We considered differences in weight scale as a possible explanation.
Our backbone was pre-trained with Muon and had weight matrices approximately five times larger in scale than those of Gemma and MiniCPM, which were pre-trained with AdamW.
We hypothesized that the shared learning rate produced larger updates relative to the smaller weights in Gemma and MiniCPM, contributing to their increases in training loss.
In other words, the learning rate appeared to be too large.

To test this explanation, we repeated supervised fine-tuning for Gemma and MiniCPM, but with a smaller peak learning rate of $1\times10^{-5}$, keeping the other training settings unchanged.
The lower rate eliminated the increase in training loss, seen in Figure~\ref{fig:lr-backbones-lowlr}.
Both backbones reached TREC DL effectiveness comparable to baseline run~1, while MiniCPM achieves a higher BEIR mean beyond the noise guideline.
These results are consistent with the proposed weight-scale explanation and suggest that learning-rate sensitivity should be considered when comparing backbones under a shared training recipe.

We find that, starting from MiniCPM, it is possible to train a pointwise reranker that is more effective than our \ourbasereranker.
However, the fact that there are ``better'' pre-trained checkpoints available to fine-tune does not detract from our core thesis.\footnote{In fact, it would be suspicious otherwise, that somehow we had stumbled onto the best pre-trained model for our task with minimal exploration of the design space.}
Our claims focus on sufficiency, that our own model checkpoint pre-trained from scratch can produce a competitive reranker, which remains true.
In other words, at least for the task of pointwise reranking, it appears that our pre-training recipe is ``good enough''.
This allows us to retain full end to end control of the training pipeline.

\subsection{The Effects of Training Data}
\label{sec:training-set-size}

\subsubsection{Experimental Setup}

It is well known, across diverse machine-learning applications, that model effectiveness depends on the size of the training dataset and the composition of the training instances.
In our next series of experiments, we attempted to isolate the effects of the data recipe.

One obvious starting point is to examine the variant datasets prepared by~\cite{thakur2025rlhn} using the same initialization and training recipe:\ RLHN-250K (247,534 instances) versus RLHN-680K (648,762 instances).
Since we applied exactly the same training recipe, training models under both data conditions for one epoch, the larger dataset increases the number of instances by a factor of 2.62 and the number of optimization steps from 1,931 to 5,060.

Note that the number of training instances is not the only difference between these variants; the source distribution also changes.
In moving from RLHN-250K to RLHN-680K, MS MARCO increases from 55.2\% to 70.2\% of the data and HotpotQA from 12.1\% to 13.0\%.
The share of Natural Questions decreases from 12.0\% to 8.9\%, although RLHN-680K contains more Natural Questions training instances than RLHN-250K.
FEVER, SCIDOCS-RR, FiQA, and ArguAna retain approximately the same instance counts and consequently account for smaller fractions of the total dataset size.

To provide a point of contrast on data composition, we also trained on 399,600 prepared instances from the Tevatron MS MARCO dataset,\footnote{\url{https://huggingface.co/datasets/Tevatron/msmarco-passage}} using the same causal backbone, LCE objective, and one-epoch training recipe.
Note that Tevatron contains only MS MARCO data, whereas RLHN-250K combines seven collections and includes LLM-assisted relabeling of false negatives~\citep{thakur2025rlhn}, so the datasets differ in terms of size, composition, and preparation.
The Tevatron and RLHN-250K datasets each provide approximately 30 negative passages per query on average.
The larger Tevatron dataset required 3,112 optimizer updates, compared with 1,931 for RLHN-250K.

\subsubsection{Results and Discussion}

\begin{table*}[t]
\startsftrows
\centering
\tableformat
\newcolumntype{R}[1]{>{\raggedleft\arraybackslash}p{#1}}
\begin{tabular}{@{}rlrR{1.25cm}R{1.25cm}@{}}
\toprule
& Training data \mbox{\hspace{1.5cm}} & Instances & TREC DL & BEIR \\
\midrule
\sftrowlabel{row:data:rlhn250k} & RLHN-250K & 247,534 & 0.641 & 0.544 \\
\sftrowlabel{row:data:rlhn680k} & RLHN-680K & 648,762 & \textbf{0.646} & \textbf{0.545} \\
\sftrowlabel{row:data:tevatron} & Tevatron MS MARCO & 399,600 & 0.618 & 0.470 \\
\bottomrule
\end{tabular}
\caption{Effects of training data (mean nDCG@10).
All rerankers have 3.22B parameters and use the same one-epoch training recipe, with differences in training data (size and source mix), GPU hardware, microbatch token budget, and resulting update count.
All configurations rerank top-100 BM25 candidates.
Row~1 is baseline run~1 of \ourbasereranker.
The highest score in each column is marked in bold.}
\label{tab:training-size-ablation}
\end{table*}

Effectiveness results are shown in Table~\ref{tab:training-size-ablation}.
Training on the larger RLHN dataset increases mean nDCG@10 on TREC DL from 0.641 to 0.646, an improvement larger than the benchmark-level noise guideline.
Gains on four of the five TREC DL collections exceed the typical collection-level noise scale, while the DL21 difference falls below the threshold.
The BEIR mean changes from 0.544 to 0.545, which we treat as experimental noise.
This average hides larger collection-specific changes, including an increase of 0.017 on TREC-News and a decrease of 0.015 on TREC-COVID.
The TREC DL gains are consistent with the increased emphasis on MS MARCO in the larger dataset, but this comparison does not isolate the effects of dataset size, source distribution, and additional optimization steps.

On TREC DL, training on RLHN-250K yields higher reranking effectiveness than training on Tevatron MS MARCO, with a difference of 0.023 in mean nDCG@10.
Both datasets included MS MARCO, but Tevatron provides more training instances and optimizer updates.
The hard-negative mining and LLM-assisted relabeling in RLHN-250K may provide higher-quality negatives with fewer false-negative labels, contributing to this difference.
The effectiveness gap is larger on BEIR, at 0.074 mean nDCG@10, consistent with the expected benefit of broader training-domain coverage.
Tevatron contains no BEIR training collections, whereas RLHN-250K includes FiQA and SCIDOCS-RR, covering two of the seven BEIR evaluation domains.

Table~\ref{tab:rlhn-checkpoints} tracks effectiveness over the course of RLHN-250K training, using the same top-100 BM25 reranking protocol as our headline results.
We see that most of the improvements in effectiveness arrive early. 
By roughly 150K queries, results on both benchmarks are close to their final scores, and the remaining instances contribute only small gains.
However, the two benchmarks appear to behave differently along the way.
Effectiveness on BEIR improved steadily throughout, but effectiveness on the TREC DL collections declined mid-training while the learning rate remained high but scores recovered as the learning rate decayed.
This pattern suggests sensitivity to the learning-rate schedule, consistent with the observations in Section~\ref{section:different-backbones}.

\begin{table*}[t]
\startsftrows
\centering
\tableformat
\newcolumntype{R}[1]{>{\raggedleft\arraybackslash}p{#1}}
\begin{tabular}{@{}lR{0.8cm}R{0.8cm}R{0.8cm}R{0.8cm}R{0.8cm}@{}}
\toprule
& \multicolumn{5}{c}{Training queries} \\
\cmidrule(l){2-6}
Benchmark & 50K & 100K & 150K & 200K & 250K \\
\midrule
TREC DL & 0.628 & 0.616 & 0.634 & 0.637 & 0.638 \\
BEIR    & 0.525 & 0.533 & 0.542 & 0.543 & 0.545 \\
\bottomrule
\end{tabular}
\vspace{4pt}
\caption{Effectiveness of baseline run~4 across RLHN-250K training checkpoints (mean nDCG@10).
All configurations rerank top-100 BM25 candidates.}
\label{tab:rlhn-checkpoints}
\end{table*}

\subsection{The Effects of Causal versus Bidirectional Modeling}
\label{sec:bidirectional-ablation}

\subsubsection{Experimental Setup}

In these experiments, we compared causal and bidirectional attention, prompt changes, and mean pooling.
All variants followed the basic experimental setup described in Section~\ref{section:sft_config}, but with changes to the starting checkpoint, attention pattern, prompt, and readout, specified below.

The two starting checkpoints were:

\begin{itemize}[leftmargin=0.5cm]

\item \ourptmodeltt (the causal variant), and

\item \ourbidirptmodeltt (the bidirectional variant).

\end{itemize}

The bidirectional variant was adapted from an earlier causal checkpoint using MNTP, as described in Section~\ref{sec:pretraining-setup}.
The two backbones differed in their final pre-training but were trained on roughly the same number of tokens.

We also tested a query--passage--query (QPQ) prompt under causal attention.
The first query conditions passage representations, while the repeated query enables the model to attend to the passage before scoring.
This placement is motivated by Jina-Reranker-v3, which places a query before and after the candidate documents~\citep{wang2025jinarerankerv3}.
This prompt design is also related to echo embeddings, in which the prompt repeats an input and extracts embeddings from the second occurrence, whose tokens can attend to the complete first occurrence under causal attention~\citep{springer2025repetition}.
In our reranker, we repeat only the same truncated query around a single passage and retain the baseline scoring function.
The short context window of our pre-trained checkpoint prohibits direct adoption of Jina's multi-document recipe, although repeating the query is feasible for our model.

Figure~\ref{fig:reranker-prompt} shows the three prompts used in supervised fine-tuning and evaluation:

\begin{itemize}[leftmargin=0.5cm]
\item \textbf{Standard:} the prompt in our headline results, tested with both causal and bidirectional attention.
\item \textbf{QPQ:} the same truncated query appears before and after the passage, with causal attention.
\item \textbf{Appended mask:} the MNTP mask token, \mytt{<|output\_end|>}, follows \mytt{Relevant:}, with bidirectional attention.
Scoring uses the preceding colon, matching MNTP's prediction position.
\end{itemize}

For both backbones, we additionally tested mean pooling with the standard prompt.
We averaged final hidden states over all non-padding tokens and applied the same two pre-trained scoring rows, fine-tuning them jointly with the backbone.
No new classification head was introduced.

\subsubsection{Results and Discussion}

\begin{table*}[t]
\startsftrows
\centering
\tableformat
\newcolumntype{R}[1]{>{\raggedleft\arraybackslash}p{#1}}
\begin{tabular}{@{}rlR{1.25cm}R{1.25cm}@{}}
\toprule
& Configuration \mbox{\hspace{1.5cm}} & TREC DL & BEIR \\
\midrule
\sftrowlabel{row:bidir:causal} & Causal, standard prompt & 0.641 & 0.544 \\
\sftrowlabel{row:bidir:qpq} & Causal, QPQ prompt & \textbf{0.647} & \textbf{0.551} \\
\sftrowlabel{row:bidir:causal-mean} & Causal, mean pooling & 0.642 & 0.543 \\
\midrule
\sftrowlabel{row:bidir:standard} & MNTP, standard prompt & 0.643 & 0.544 \\
\sftrowlabel{row:bidir:mask} & MNTP, appended mask & 0.643 & 0.543 \\
\sftrowlabel{row:bidir:mean} & MNTP, mean pooling & 0.638 & 0.538 \\
\bottomrule
\end{tabular}
\caption{Effects of causal versus bidirectional modeling (mean nDCG@10).
All rerankers have 3.22B parameters and use the same training recipe, with differences in pre-trained initialization, attention, prompt, and readout.
All configurations rerank top-100 BM25 candidates.
Row~1 is baseline run~1 of \ourbasereranker.
The highest score in each column is marked in bold.}
\label{tab:bidirectional-ablation}
\end{table*}

Table~\ref{tab:bidirectional-ablation} compares the effects of attention pattern, prompt, and pooling on ranking effectiveness.
With the standard prompt, the causal baseline and the bidirectional reranker initialized from the MNTP checkpoint achieve comparable effectiveness, with differences on the scale of run-to-run noise (rows~\ref{row:bidir:causal} and~\ref{row:bidir:standard}).
Appending the mask token produces benchmark means comparable to the standard bidirectional prompt (rows~\ref{row:bidir:standard} and~\ref{row:bidir:mask}).

Compared to the standard causal prompt (row~\ref{row:bidir:causal}), the causal QPQ prompt (row~\ref{row:bidir:qpq}) improves mean nDCG@10 from 0.641 to 0.647 on TREC DL and from 0.544 to 0.551 on BEIR, with both changes larger than the noise guideline.
Repetition may help by giving the model query representations that have been conditioned by the candidate document before the final scoring position, without requiring any changes to its pre-trained causal attention pattern.
This is a possible explanation for the observed effectiveness improvement.

To understand why bidirectional adaptation yields ranking effectiveness comparable to the causal baseline, we tested whether the adapted rerankers attended to later tokens.
We selected DL19, DL20, and SciFact to cover both TREC DL and BEIR under the in-domain classification used in Section~\ref{section:headline:experimental-setup}.
Their small query sets (43, 54, and 300 queries, respectively) kept the diagnostic evaluation inexpensive.
With model weights fixed, blocking attention to later tokens reduced mean nDCG@10 across these three collections from 0.753 to 0.700 for the standard prompt and from 0.750 to 0.688 for the mask-appended prompt.

These decreases show that the adapted models rely on bidirectional attention.
However, the baseline trained and evaluated with causal attention achieves comparable ranking effectiveness, suggesting that the model has learned alternative ways to combine query and passage information.

For the bidirectional model, replacing the single-position readout (row~\ref{row:bidir:standard}) with mean pooling (row~\ref{row:bidir:mean}) reduces mean nDCG@10 from 0.643 to 0.638 on TREC DL and from 0.544 to 0.538 on BEIR, with both declines larger than the noise guideline.
However, for the causal model, the differences between the last non-padded token and mean pooling readout are within the noise guideline, suggesting comparable effectiveness (rows~\ref{row:bidir:causal} and~\ref{row:bidir:causal-mean}).
Together, these results show that the adapted models {\it do} make use of bidirectional attention, but without a clear reranking advantage over the causal baseline.
Similarly, replacing their single-position readout with mean pooling did not appear to improve reranking effectiveness.

A possible explanation is that MNTP improves masked-token prediction without sufficiently improving the representations used for relevance scoring, and hence has minimal impact on reranking.
We leave pre-training a BERT-style bidirectional model~\citep{devlin2019bert} from scratch and exploring other pooling methods to future work.

\subsection{The Effects of Self-Filtering and Self-Distillation}
\label{sec:self-filtering-distillation}

\subsubsection{Experimental Setup}

In these experiments, we compared the baseline trained on RLHN-250K with two variants:\ additional self-filtering of the training data and self-distillation on the self-filtered data.
The first comparison examined the effect of filtering; the second examined the added value of score supervision from the trained baseline checkpoint.

We used the model checkpoint from baseline run 1 to identify potentially noisy positives in RLHN-250K.
For each training query, we scored all positive and negative passages and removed a positive if more than $\max(9,0.375N)$ negatives received a higher score, where $N$ is the total number of negatives available for that instance before sampling.
Queries with no remaining positive relevance judgments were then discarded.
We did not apply this procedure to ArguAna, which was one of the seven training sources in RLHN-250K~\citep{thakur2025rlhn}, because its task requires retrieving counterarguments, which may not align with the model's general relevance judgments.

This filtering procedure removed 14,887 positive passages (2.6\%) and 1,484 queries (0.6\%), leaving 246,050 queries in the dataset.
We then trained a new reranker on the retained data using the same LCE objective and training recipe as our baseline run~1.
The checkpoint from baseline run~1 supplied the filtering decisions, while the new reranker was initialized from the original pre-trained checkpoint.

We also explored whether the baseline checkpoint can provide useful supervision beyond data filtering.
One common ranking-distillation objective is Margin-MSE~\citep{hofstatter2020marginmse}, which minimizes the squared difference between teacher and student score margins for pairs of passages.
We did not select this objective, but instead used Kullback--Leibler (KL) divergence to match LCE's groupwise softmax formulation:\ LCE supervises the documents with positive labels, while KL divergence transfers the teacher's relative preferences over the sampled negatives to the student.
For a fixed teacher, minimizing KL divergence is equivalent to minimizing cross-entropy with soft targets.
This motivates our objective formulation here, without making any empirical claims about effectiveness compared to Margin-MSE.

Putting everything together, on the self-filtered data, the distillation objective is:
\begin{align}
p_T &= \operatorname{softmax}(\mathbf{s}^{T}_{-}/\tau_{\mathrm{KD}}), &
p_S &= \operatorname{softmax}(\mathbf{s}^{S}_{-}/\tau_{\mathrm{KD}}),\\
\mathcal{L} &= \mathcal{L}_{\mathrm{LCE}}+0.3\,\tau_{\mathrm{KD}}^2 D_{\mathrm{KL}}(p_T\|p_S).
\end{align}
Both teacher and student use $\tau_{\mathrm{KD}}=2$, while LCE retains $\tau=1$.
The factor $\tau_{\mathrm{KD}}^2$ compensates for the temperature's effect on the gradient scale.
Here, $\mathbf{s}^{T}_{-}$ and $\mathbf{s}^{S}_{-}$ are teacher and student scores over the $K$ negatives.
The positive documents are excluded from distillation, leaving their supervision to LCE.
The teacher is fixed; the student has the same architecture and starts from the original pre-trained initialization.
Distillation was not performed for ArguAna.

\subsubsection{Results and Discussion}

\begin{table*}[t]
\startsftrows
\centering
\tableformat
\newcolumntype{R}[1]{>{\raggedleft\arraybackslash}p{#1}}
\begin{tabular}{@{}rlR{1.25cm}R{1.25cm}@{}}
\toprule
& Training configuration \mbox{\hspace{2cm}} & TREC DL & BEIR \\
\midrule
\sftrowlabel{row:filter:baseline} & LCE & \textbf{0.641} & 0.544 \\
\sftrowlabel{row:filter:self} & LCE + self-filtering & \textbf{0.641} & \textbf{0.548} \\
\sftrowlabel{row:filter:kd} & LCE + self-filtering + self-distillation & 0.640 & 0.547 \\
\bottomrule
\end{tabular}
\caption{Effects of self-filtering and self-distillation (mean nDCG@10).
All rerankers have 3.22B parameters and use the same pre-trained initialization and training recipe, except for self-filtering, the additional distillation loss, and the resulting update count.
All configurations rerank top-100 BM25 candidates.
Row~1 is baseline run~1 of \ourbasereranker.
The highest score in each column is marked in bold (including ties).}
\label{tab:training-ablation}
\end{table*}

Experimental results are shown in Table~\ref{tab:training-ablation}.
We see that self-filtering improves mean BEIR nDCG@10 from 0.544 to 0.548, beyond the benchmark-level noise guideline, while TREC DL effectiveness remains within run-to-run noise.
The largest observed gain was on TREC-News, increasing from 0.510 to 0.529 and exceeding the typical collection-level noise threshold.
Adding self-distillation with matched temperatures produces benchmark means comparable to self-filtering alone; both differences are within the noise guideline.
Collection-level changes appear mixed:\ FiQA increased by 0.004, but DL23 and SciFact decreased by 0.007 and 0.008, respectively.
These results suggest that the trained checkpoint can provide useful filtering decisions, but that score supervision from a teacher of the same capability appears to offer no clear additional benefits in reranking effectiveness.

A common alternative, naturally, is to distill supervision from a stronger teacher.
In line with our focus on sovereign AI, we opted for self-filtering and self-distillation using variants of our own models to avoid an additional external teacher over which we may not have control.
We leave alternative formulations of distillation for future work.

\subsection{The Effects of Different Loss Functions}
\label{sec:ce-loss-ablation}

\subsubsection{Experimental Setup}

In these final experiments, we compared LCE with pointwise cross-entropy (CE), implemented as answer-token prediction.
The CE experiment used the same sampled positive and negative query--document pairs and the same optimizer-step plan as LCE.
This setting also matched the pre-trained initialization, standard prompt, optimization settings, random seed, training precision, and input-length limits.

With the CE loss function, we trained the model to take each pair as input to predict a single \mytt{\textvisiblespace{}true} or \mytt{\textvisiblespace{}false} answer token.
This approach retains the full pre-trained language-model output layer and averages answer-token cross-entropy over the pairs, with normalization over the entire vocabulary.
With LCE, the model instead retains only the two scoring rows and averages contrastive losses over query groups.
This comparison therefore evaluates standard supervised fine-tuning against our LCE recipe under matched training exposure. 

\subsubsection{Results and Discussion}

\begin{table*}[t]
\startsftrows
\centering
\tableformat
\newcolumntype{R}[1]{>{\raggedleft\arraybackslash}p{#1}}
\begin{tabular}{@{}rlR{1.25cm}R{1.25cm}@{}}
\toprule
& Objective \mbox{\hspace{1.0cm}} & TREC DL & BEIR \\
\midrule
\sftrowlabel{row:loss:lce} & LCE & \textbf{0.641} & \textbf{0.544} \\
\sftrowlabel{row:loss:ce} & Pointwise CE & 0.628 & 0.528 \\
\bottomrule
\end{tabular}
\caption{The effects of different loss functions (mean nDCG@10).
Both variants start from the same 3.29B-parameter pre-trained checkpoint; the rerankers retain 3.22B parameters for LCE and 3.29B for CE.
All configurations rerank top-100 BM25 candidates.
Row~1 is baseline run~1 of \ourbasereranker.
The highest score in each column is marked in bold.}
\label{tab:loss-ablation}
\end{table*}

Experimental results are shown in Table~\ref{tab:loss-ablation}.
We see that LCE achieves higher mean nDCG@10 than CE on both benchmarks:\ 0.641 versus 0.628 on TREC DL and 0.544 versus 0.528 on BEIR.
The absolute improvements of 0.013 and 0.016, respectively, are larger than the run-to-run noise guideline.
These results support LCE for our decoder-based reranker under the tested recipe and are consistent with the encoder-based findings of \citet{pradeep2022squeezing}.

\section{Related Work}
\label{section:related-work}

\subsection{AI Sovereignty}

The idea of AI sovereignty is not new.
Canada's national AI strategy from 2017 already emphasized retaining domestic talent and economic benefits from AI.\footnote{\href{https://www.budget.canada.ca/2017/docs/plan/chap-01-en.html}{Canada's Budget 2017: Skills, Innovation and Middle Class Jobs.}}
In a speech dating to March 2018, French President Emmanuel Macron linked AI infrastructure and development to French sovereignty and retaining economic value in France and Europe.\footnote{\href{https://www.elysee.fr/emmanuel-macron/2018/03/29/discours-du-president-de-la-republique-sur-lintelligence-artificielle}{Emmanuel Macron's speech on AI, March 29, 2018.}}
The concept has been steadily gaining prominence:\
In February 2024, NVIDIA founder and CEO Jensen Huang asserted at the World Governments Summit in Dubai that every country needs to own the production of their own intelligence.\footnote{\href{https://blogs.nvidia.com/blog/world-governments-summit/}{Jensen Huang on sovereign AI at the World Governments Summit, February 12, 2024.}}
Sovereign AI has also attracted substantial attention from major consulting firms such as Deloitte, McKinsey, Accenture, and BCG, all of which have opined on various aspects of this complex topic.
The Linux Foundation~\citep{Gerosa_etal_2025} identified open source as a foundation for sovereignty through transparency, adaptability, and reducing dependence on individual vendors.
Although initially articulated through the lens of national concerns, sovereignty has increasingly been discussed at the level of individual organizations.
Recent episodes concerning access to frontier models from Anthropic (discussed in Section~\ref{section:sovereign:organizational}) have thrust questions about AI sovereignty into the spotlight.

Alex Karp, the CEO of Palantir, has prominently advocated for AI sovereignty in recent months, emphasizing the importance of organizations retaining control over their institutional knowledge and their ``means of production''.
These arguments were more precisely laid out in August 2026,\footnote{\href{https://www.palantir.com/q2-2026-letter/en}{Alex Karp's Q2 2026 letter to shareholders, August 3, 2026.}} but they can be traced back to public interviews in July.
He warns of model providers misappropriating customers' business data and capturing their ``alpha'' (an organization's distinctive knowledge and capabilities that give it a competitive advantage).
Another concern he raised is the cost associated with LLMs and organizations not necessarily deriving value from token spend.
LLM inference costs provide another rationale for the pursuit of sovereign AI:\ any organization dependent on external model providers will have limited control over pricing charged by third parties.
Thus, as part of sovereignty efforts, many organizations are exploring whether smaller, and therefore cheaper, task-specific models might be just as effective as frontier models (more discussion below).

In a white paper from July 2026 to complement many of Karp's arguments, \cite{Palantir_2026} argues for sovereignty from the perspective of an organization owning the value created through AI and the freedom to pursue new opportunities.
The choices that an organization makes with respect to AI model usage can narrow or broaden these economic rights.
The report laid out recommendations spanning models, compute, and operational control.
It argues that zero data retention alone is insufficient:\ organizations must also preserve the ability to switch providers and retain the knowledge and improvements generated through their use of AI.

While perhaps not as outspoken as Alex Karp, Microsoft CEO Satya Nadella in July 2026 made similar arguments about AI sovereignty in what he calls the ``Reverse Information Paradox''.\footnote{\url{https://x.com/satyanadella/status/2076323181154230284}}
He asserted that, with AI, organizations pay for intelligence twice, once with money, and again with something even more valuable:\ proprietary knowledge that an organization must reveal to make that intelligence useful.
Over time, model providers learn about their customers (via prompts, confidential information, agent tools, human feedback, usage statistics, etc.), making it imperative for organizations to establish a trust boundary to prevent misappropriation.
Nadella actually cites Karp, saying that organizations ``want to know they own the means of production, and it's not being transferred to someone else''.

Our work is very much inspired by this line of thinking, and we share concerns about institutional autonomy.
To this discussion, we additionally contribute two less frequently mentioned issues from the academic perspective:\ the important role of transparency in developing a rigorous understanding of model behavior and the valuable educational opportunities that arise from the ability to build fully open and sovereign models (Section~\ref{section:sovereign:academic}).
From a technical perspective, we focus specifically on the transparency and reproducibility of model development, including access to the data and recipes needed to independently build various artifacts.
We make available everything necessary for an organization to reproduce our results, to inspect each step in the pipeline, and to make any modifications as necessary.
Our contribution is to demonstrate what can be achieved with limited computational resources in a fully open and sovereign manner.

\subsection{Open Models and Related Efforts}

We compare the goals of Project Greenhouse and our current contributions to a representative selection of other open models and related efforts.

\paragraph{OLMo and T\"{u}lu.}
Building on earlier open efforts such as BLOOM~\citep{bloom} and Pythia~\citep{pythia}, AI2 announced the OLMo (Open Language Model) project in May 2023.\footnote{\href{https://medium.com/ai2-blog/announcing-ai2-olmo-an-open-language-model-made-by-scientists-for-scientists-ab761e4e9b76}{Announcing AI2 OLMo, an Open Language Model Made by Scientists, for Scientists.}}
The goal was to develop fully open language models to benefit the research community by providing access for scientific study and education around all aspects of model creation.
AI2 accomplished this by releasing not only model checkpoints, but also training data, code, and evaluation resources.
The first OLMo models were released in February 2024~\citep{groeneveld-etal-2024-olmo}, followed by OLMo 2 in November 2024~\citep{olmo2} and Olmo 3 in November 2025~\citep{olmo3}.
This seminal line of work has provided an important reference point for open research in building language models and establishes an important foundation for our approach:\ transparency throughout model development enables independent experimentation, reproducibility, and rigorous understanding.

% https://allenai.org/blog/olmo2
% https://allenai.org/blog/olmo3

Nevertheless, there are important differences between the OLMo project and our efforts:\ 
OLMo aspires to be a family of state-of-the-art models with a diverse set of capabilities, including long-context reasoning, function calling, coding, instruction following, general chat, and knowledge recall.
In contrast, our ambitions are much more tightly scoped, focused only on agentic search.
Due to its frontier aspirations, OLMo required far more compute than typically available to academic institutions:
The developers reported approximately 56 days from the start of training to the evaluation of Olmo 3 Think 32B on a cluster with 1,024 H100 GPUs~\citep{olmo3}.
The model was pre-trained on roughly 6T tokens, approximately 20$\times$ larger than our corpus.

Despite this transparency, reproducing the larger OLMo models from scratch remains beyond the resources of most individual research groups.
In contrast, our reranker models are far more restricted in capability but can be reproduced using only a handful of GPUs.
We therefore view our efforts as complementary:\ alongside transparency in how models are developed, we emphasize making end-to-end experimentation practical under modest compute budgets.
This accessibility matters because researchers need to do more than reproduce a single checkpoint; they must also be able to modify the data, test alternative recipes, and run controlled ablations.

T\"{u}lu is a complementary project~\citep{tulu}, also by AI2, that focuses on open post-training methods for existing models.
The effort partially overlaps with the OLMo project and its recipes informed the development of instruction-following OLMo models.
For example, T\"{u}lu 3 builds on Llama 3.1 base models, with releases that include data, code, and recipes used for post-training.
DR Tulu~\citep{ShaoRulin_etal_ICML2026} is an agentic search model directly trained for open-ended, long-form deep research.
Starting from Qwen3-8B, the authors performed supervised fine-tuning on research trajectories, including examples generated from GPT-5, followed by reinforcement learning with evolving rubrics (RLER).
These rubrics incorporate retrieved evidence and are updated using the model's rollouts, allowing the evaluation criteria to adapt to newly discovered information.
The project released model weights, training data, code, and search-agent infrastructure.

While DR Tulu shares some of the same end goals as our efforts, there are substantial differences in scope and approach.
Project Greenhouse aspires to tackle agentic search in its entirety (which includes deep research), but our approach is more incremental:\ we plan to develop and evaluate specific capabilities before integrating them into more complex systems.
Our first milestone is a pointwise reranker, which is far less ambitious in scope than DR Tulu.
However, we require that the resulting models be fully open and sovereign under our definition.
Although DR Tulu makes its agent-training pipeline available, it begins with an open-weight backbone whose pre-training data and recipes are not fully available.
Thus, it does not qualify as fully open and sovereign under our definition.
We will not repeat arguments for why we feel that open-weight models are not enough, but see Section~\ref{section:sovereign:organizational} for the organizational perspective and Section~\ref{section:sovereign:academic} for the academic perspective.

\paragraph{Marin.}
The purpose of the Marin project is the discovery and promulgation of the ``process knowledge'' necessary to create frontier-level artificial intelligence.\footnote{\href{https://openathena.ai/blog/open-development-of-frontier-ai/}{Open Development of Frontier AI.}}
It originated at Stanford University in 2024, but the Marin models are now primarily developed by Open Athena, a nonprofit that accelerates academia with capabilities from the AI frontier, with community contributions.\footnote{\href{https://marin.community/}{Marin project website.}}
According to the Marin website, ``open'' means that everything is shared:\ all discussions, all code, including models, infrastructure, etc., model weights, the sources of all the data in the training corpus, the results of successful and unsuccessful experiments, and tools that make it easy to learn from this effort.
As of October 2026, the project is training a mixture-of-experts model with 535 billion parameters total, 23 billion parameters active, on 18 trillion tokens of data.

Marin articulates a laudable vision of model development, but Project Greenhouse differs in several important respects.
Whereas Marin explicitly aims for frontier-level intelligence, we have far more tightly scoped ambitions focused on agentic search.
Aiming for the frontier requires substantial centralized computational resources; for example, the hero run referenced above is being trained on 11$\times$ GB200 NVL72 racks.
In contrast, our approach allows for more decentralization, as most individual experiments described in this paper can be reproduced with only one or two GPUs.
Thus, we view our efforts as complementary, but sharing the same goal of expanding public knowledge.

\paragraph{Thomson.}
In August 2026, \citet{ChenShengzhuang:2608.27147:2026} contributed to the sovereign AI discussion through Thomson, a new general-purpose frontier model family trained with an enhanced focus on high-stakes professional work.
Although not technically open by our definition, discussions of sovereignty figured prominently in the narrative around the model and its associated technical report, making a focused discussion of the Thomson effort worthwhile.
The research team demonstrated that frontier performance is achievable by a wide range of institutions through continual learning on existing open-weight models (specifically, Qwen3.6-35B and Qwen3.5-397B).
Model weights for the small 35B variant of Thomson are publicly available.

While we share their emphasis on institutional autonomy and on making model development transparent, there are several important differences between the Thomson effort and Project Greenhouse.
Again, like Marin and OLMo discussed above, Thomson aims for the frontier, which we do not.
However, instead of starting ``from scratch'', i.e., from pre-training, Thomson starts from Qwen open-weight models.
Thus, the model does not meet our definition of fully open and sovereign as upstream details remain unavailable.
In working with Qwen as a starting point, the Thomson researchers expended considerable effort in ``value re-alignment'' before they could proceed with continued pre-training to focus on desired capabilities.
Furthermore, the resources required for their efforts are beyond most organizations:\ the model family was ``developed by a technical team not exceeding three dozen engineers and scientists on a modest compute cluster with no more than 368 B200 GPUs available at any stage of experimentation''.
While the project represents an important example of a viable alternative approach, its resource requirements remain beyond the budget of most academic institutions.

\paragraph{Other open models.}
The discussions above focus on prominent, high-profile efforts, but researchers have released many open models over the years. 
Earlier contributions to transparent model development include LLM360's Amber and CrystalCoder in 2023~\citep{llm360}, followed by MAP-Neo~\citep{map-neo} and DCLM~\citep{dclm} in 2024, which made training data and development artifacts available alongside model weights.

Subsequent efforts have pursued several complementary directions:\
SmolLM~\citep{smollm2} emphasizes small models and transparent training recipes, while Comma, introduced with Common Pile in 2025~\citep{common-pile}, emphasizes public-domain and openly licensed training data. 
Continuing development is evident in LLM360's subsequent K2 models~\citep{k2}, the expanding Pleias family~\citep{pleias}, including Baguettotron, and multilingual initiatives such as Apertus~\citep{apertus} and EuroLLM~\citep{eurollm}, with further releases and technical reports appearing in 2025--2026.
Most recently, Gensyn's Open-1B~\citep{open-1b}, announced in September 2026, explores independently auditable training through the release of data, checkpoints, and a framework for replaying individual training steps.\footnote{However, its dependence on an unreleased operator-library prevents us from considering the effort fully open.}

\paragraph{Qwen and other open-weight models.}
While there are many open-weight models (too numerous to enumerate), the Qwen family of models~\citep{yang-etal-2024-qwen2,qwen3embedding,qwen-2026-qwen35,qwen38} deserves special recognition as the starting point most commonly used by information retrieval researchers.
However, as we have argued throughout this report, the availability of model weights is insufficient from both the perspective of organizational control and academic transparency.

\paragraph{Other organization- and task-specific models.}
In part as a reaction to the sovereignty considerations discussed throughout this report, 
organizations are increasingly adapting models to their own tasks, seeking better performance, lower costs, and greater control over model behavior.
These efforts share the common theme of attempting to reduce dependence on proprietary frontier models by substituting open-weight backbones.

A blog post in June 2026 shared details on a collaboration between Thinking Machines and Bridgewater AIA Labs to build small models that capture human expertise on six information filtering tasks.\footnote{\href{https://thinkingmachines.ai/news/learning-to-replicate-expert-judgment-in-financial-tasks/}{Learning to Replicate Expert Judgment in Financial Tasks}}
Not surprisingly, they demonstrated that small custom task-specific models, fine-tuned on high-quality proprietary datasets labeled by expert investors, can exceed frontier performance.

The legal AI startup Harvey introduced in August 2026 their Tenet model,\footnote{\href{https://www.harvey.ai/blog/post-training-update-harvey-tenet}{Harvey Tenet Research Preview}} post-trained from a Kimi K3 base in collaboration with Fireworks research for long-horizon legal work.
The blog post indicated promising results for both performance and cost-efficiency, and significant gains on complex, document-intensive tasks like M\&A Diligence.
The company followed up in September 2026 with another blog post\footnote{\href{https://www.harvey.ai/blog/post-training-rlm-agents-for-m-and-a-diligence}{Post-Training RLM Agents for End-to-End M\&A Diligence}} about training a recursive language model~\citep{Zhang:2512.24601:2025} harness in collaboration with Baseten.

Shopify provides another example of an organization training small models for specialized use cases, as disclosed by their CEO in September 2026.\footnote{\url{https://x.com/tobi/status/2094808564355191249}}
Together, all of the instances discussed above represent public case studies of organizations grappling with AI sovereignty---control over their AI model usage from multiple perspectives (performance, cost, dependence, etc.).

\section{Limitations}

Don't like this work?
Or perhaps there are too many limitations?
We'll pre-empt criticisms in the form of anticipated objections in an attempt to self-own:

\paragraph{Little novelty.}
Yes, the often-used criticism and the favorite of reviewer 2.
Project Greenhouse lays out an overall vision and in this report we deliver on the first project milestone.
We concede that there is nothing particularly novel about each of our steps, which build on well-known methods from the literature, except that to our knowledge, we are the first to have put everything together in this way.

\paragraph{Reranking is too easy.}
Yes, it could indeed be the case that reranking is too easy for {\it any} modern LLM, and this is why we are able to demonstrate competitive effectiveness with such a simple two-step recipe.
Even if true, this finding is entirely orthogonal to the claims we make in this paper, which is focused on building fully open and sovereign models.
We believe that our goals are possible with limited computational resources, which dictates the model sizes that we can practically explore.
The fact that competitive rerankers are possible using small models is not inconsistent with our core thesis.
We will follow up with additional explorations to better characterize the interplay between model size and task difficulty.

\paragraph{Agentic search is too hard.}
While we have demonstrated the viability of fully open and sovereign models for reranking, it might be the case that the full suite of agentic capabilities is beyond what can be achieved in a fully open and sovereign manner with limited computational resources.\footnote{Although evidence from \cite{ShaoRulin_etal_ICML2026} might suggest otherwise.}
In other words, a competitive reranker does not establish a path to a competitive agent.

Note that as discussed in Section~\ref{section:agentic-search}, we do not prescribe any particular architecture, and it could be the case that non-open and non-sovereign models are still needed for at least some aspects of agentic search.
For example, a frontier or near-frontier model might still be necessary for overall coordination, while delegating other aspects of search to smaller open and sovereign models.
If this is the case, our core thesis will not be fully validated.
This may very well be true, but we won't know unless we try---and along the way, we're sure much will be learned.
We'll try it and report back.

\paragraph{You've just moved the sovereignty boundary upstream.}
Based on our model of LLM training as hypergraph traversal (Section~\ref{section:sovereign}), we defined sovereignty in terms of tracing provenance to publicly available datasets such as ClimbMix~\citep{diao-etal-2025-nemotron} and RLHN-250K~\citep{thakur2025rlhn}.
As we conceded, this is based on judgment, but we might be criticized for drawing arbitrary boundaries.
For example, datasets can contain undisclosed filtering, synthetic generation, annotation decisions, and dependencies on other models.

A skeptical reader could ask why treating a dataset as a trusted starting point is acceptable while treating a model checkpoint that way is not.
We argue that this distinction is defensible because access to individual training instances permits inspection and removal, which is not possible with access to model weights alone.
Perhaps a refinement of our sovereignty claim is the necessity of explicitly declaring provenance boundaries.

\paragraph{You could have just started with an existing fully open backbone.}
While we have argued that open-weight models are not sufficient to achieve our goals, these arguments do not necessitate our specific development path, specifically, pre-training our own backbone from scratch (Section~\ref{section:pt}).
Fine-tuning an existing small, fully open backbone might provide the same upstream transparency and perhaps better effectiveness or lower development cost.

This may perhaps be the case, but we make several important counterpoints:
Our claims primarily center around providing an ``existence proof'' of what's possible (i.e., sufficiency); we never claimed that our path is uniquely necessary.
From the perspective of reproducibility and pedagogy, however, it was desirable for us to pre-train a backbone from scratch:\ the backbone described in Section~\ref{section:pt} is simply what we ended up with.
Indeed, it would be interesting future work to extend the experiments in Section~\ref{section:different-backbones} to include other open backbones.

\paragraph{You didn't try this \ldots}
Yes, indeed.
There are many data and training conditions that we did not have time to explore and document in this report.
We attempted to strike a balance between velocity and achieving our milestone objective, separating the experiments that are necessary to support our claims from the ``nice to have'' interesting extensions.
Of course, more experiments can improve our understanding of LLM behavior.
However, we believe that our headline result provides a small step in demonstrating the veracity of our core thesis.
We invite the community to contribute to our efforts and believe that our hypergraph model of LLM training provides a lightweight mechanism for coordinating experimental activity across diverse research groups around the world.

\paragraph{You're overfitting.}
Perhaps.
While some of our experiments can be considered out of domain (see Section~\ref{section:headline:experimental-setup}), they are not as thorough as we would have liked, and this is an admitted weakness of our present results.
However, this limitation does not detract from the central claims in this report about model openness and sovereignty.
The standard tricks that the community has developed to promote generalization and guard against overfitting remain applicable to our training pipeline, and there is no reason why we can't also adopt those techniques.
We will follow up in subsequent work to explicitly explore this question of out-of-domain generalization and report back.

\paragraph{You'll never win.}
A provider of proprietary or non-sovereign models can always distill down to whatever model size and scope of capabilities that we're targeting.
In that way, we will always be dominated along the Pareto frontier and will never achieve ``best in class'' status.
This is likely true, but our objective was never to ``win''.
Our thesis centers on demonstrating capabilities in a fully open and sovereign manner, with the bar of achieving ``competitive'' effectiveness (as opposed to achieving the state of the art).
In this way, our models will support something that no proprietary or non-sovereign model can:\ transparency and the rigorous understanding that comes along with it.

\section{Conclusion}

Project Greenhouse intends to be an ongoing, long-term project and this report lays out our goals at a high level, including arguments on why we believe our efforts are important and potentially impactful.
Our goals are ambitious but, we feel, achievable, and here we deliver the first project milestone.
We describe how to build a competitive pointwise decoder-only reranker with a simple two-step recipe comprising pre-training from scratch and supervised fine-tuning, starting only from commonly available datasets.
As we do not rely on existing open-weight backbones from third parties, we are fully in control of model training, from end to end.
The modest computational requirements of our experiments mean that it is feasible for others to reproduce and to build on our results.

The path forward is clear for us:\ having laid out our vision and delivered the first project milestone, we'll simply continue and do what we said we would.
We extend an invitation for other like-minded academic research groups, organizations, and independent researchers to join us on our journey.
Otherwise, stay tuned for the next installment!

\begin{ack}
This research was supported in part by the Natural Sciences and Engineering Research Council of Canada (NSERC) and GPUs from the Digital Research Alliance of Canada.
\end{ack}

\bibliographystyle{plainnat}
\bibliography{greenhouse}

\clearpage

\appendix

\section{Listwise Reranking Prompt}
\label{app:listwise-prompt}

Figure~\ref{fig:listwise-prompt} reproduces the RankLLM prompt used for the Gemma-4, Qwen3.8, and GPT-6.1 Sol listwise reranker baselines in Section~\ref{section:headline}, taken from~\cite{sharifymoghaddam2025rankllm}.
We used the \mytt{qwen\_non\_thinking\_template.yaml} template from RankLLM revision \mytt{443adf3}.\footnote{\href{https://github.com/castorini/rank_llm/blob/443adf3306102f11b85945d2a277fb6aa10ea0aa/src/rank_llm/rerank/prompt_templates/qwen_non_thinking_template.yaml}{RankLLM prompt source (revision \mytt{443adf3}).}}

Qwen3.8 used the system message as shown.
For Gemma-4, we omitted the Qwen-specific \mytt{/no\_think} line and disabled thinking through the model's chat template. 
For GPT-6.1 Sol, we implemented our listwise reranker using Pi, concatenating the RankLLM template's system message prefix (omitting the \mytt{/no\_think} line), passage entries, and suffix into a single user message and invoking Pi with \mytt{\-\-thinking minimal}.
In Pi, we disabled all tools, extensions, skills, and automatic context-file loading, so the model does not have access to tools for reading files, executing commands, or accessing external resources.
For each sliding window, we substituted its passage count for \mytt{\{num\}}, the query for \mytt{\{query\}}, and each passage's identifier and text for \mytt{\{rank\}} and \mytt{\{candidate\}}.
The user message concatenates the prefix, one body entry per passage, and the suffix. 

\begin{figure}[!htbp]
\centering
\textbf{System message}\par
\begin{mdframed}[roundcorner=6pt,linewidth=1pt,innerleftmargin=6pt,innerrightmargin=6pt,innertopmargin=5pt,innerbottommargin=5pt]
\begin{Verbatim}[fontsize=\footnotesize,breaklines=true,breakanywhere=false,breaksymbolleft={}]
/no_think

You are RankLLM, an intelligent assistant that can rank passages based on their relevancy to the query.
Given a query and a list of passages, your task is to re-rank these passages based on their relevance to the query.

Please perform the following steps:
1. **Understand the Query**: First, carefully read and understand the user's query to identify the core information need.
2. **Analyze Each Passage**: For each passage, critically evaluate its content and determine how well it addresses the query. Consider factors like:
    - Directness of the answer
    - Completeness of the information
    - Presence of supporting evidence or details
    - Absence of irrelevant or distracting information
3. **Compare and Contrast**: Compare the passages against each other. Identify which passages are more relevant and why. Note any subtle differences in relevance.
4. **Assign Ranks**: Based on your analysis, assign a unique rank to each passage, starting from 1 for the most relevant.

**Output Format:**
Your final output must be a list of ranks, corresponding to the original order of the passages. For example, if there are 3 passages, and you decide the second passage is most relevant, the first is second most relevant, and the third is least relevant, your output should be:
[2] > [1] > [3]

No other text or explanation should be present in the final output, only the list of ranks.
\end{Verbatim}
\end{mdframed}
\textbf{User message}\par
\begin{mdframed}[roundcorner=6pt,linewidth=1pt,innerleftmargin=6pt,innerrightmargin=6pt,innertopmargin=5pt,innerbottommargin=5pt]
\begin{Verbatim}[fontsize=\footnotesize,breaklines=true,breakanywhere=false,breaksymbolleft={}]
I will provide you with {num} passages, each indicated by a numerical identifier []. Rank the passages based on their relevance to the search query: {query}.

[{rank}] {candidate}
...

Search Query: {query}.
Rank the {num} passages above based on their relevance to the search query. All the passages should be included and listed using identifiers, in descending order of relevance. The output format should be [] > [], e.g., [2] > [1], Answer concisely and directly and only respond with the ranking results, do not say any word or explain.
\end{Verbatim}
\end{mdframed}
\caption{Listwise prompt used for zero-shot reranking with Qwen3.8, Gemma-4, and GPT-6.1 Sol.
The ellipsis denotes repeated passage entries; Gemma-4 and GPT-6.1 Sol omitted the opening \mytt{/no\_think} directive.}
\label{fig:listwise-prompt}
\end{figure}

\section{Jev Shared-Rubric Prompt}
\label{app:jev-rubric-prompt}

Figure~\ref{fig:jev-rubric-prompt} shows the shared-rubric request used for Jev's TREC DL results in Section~\ref{section:headline}.
We used the \mytt{jev-latest} model from TypeSafe AI and supplied the query, all top-100 BM25 candidates in retrieval order, and the four-level TREC relevance-grade rubric in a single request, with one \mytt{score} question per passage.

We ranked passages by the API's returned expected grade, $s_i=\sum_{g=0}^{3}g\,p_i(g)$, where $p_i(g)$ is the probability assigned to relevance grade $g$ for passage $i$.
Score ties retained the original BM25 order.
We used the same query and passage truncation limits as \ourbasereranker, 128 and 512 tokens, respectively, with \mytt{cl100k\_base} tokenization.

\begin{figure}[!htbp]
\centering
\begin{mdframed}[roundcorner=6pt,linewidth=1pt,innerleftmargin=6pt,innerrightmargin=6pt,innertopmargin=5pt,innerbottommargin=5pt]
\begin{Verbatim}[fontsize=\footnotesize,breaklines=true,breakanywhere=false,breaksymbolleft={}]
{
  "state": {
    "query": "<query>",
    "passages": {"P1": "<passage 1>", "P2": "<passage 2>"},
    "relevance_levels": {
      "Irrelevant": "Irrelevant: the passage has nothing to do with the query.",
      "Related": "Related: the passage seems related to the query but does not answer it.",
      "Highly relevant": "Highly relevant: the passage has some answer for the query, but the answer may be a bit unclear, or hidden amongst extraneous information.",
      "Perfectly relevant": "Perfectly relevant: the passage is dedicated to the query and contains the exact answer."
    }
  },
  "questions": {
    "P1": {
      "type": "score",
      "instructions": "How relevant is passage P1 to the query?",
      "criteria": ["Irrelevant", "Related", "Highly relevant", "Perfectly relevant"]
    },
    "P2": {
      "type": "score",
      "instructions": "How relevant is passage P2 to the query?",
      "criteria": ["Irrelevant", "Related", "Highly relevant", "Perfectly relevant"]
    }
  }
}
\end{Verbatim}
\end{mdframed}
\caption{Jev reranking prompt using the four-level TREC relevance-grade rubric at evaluation time.
Two passages illustrate the request structure; each evaluated request contained all 100 BM25 candidates and one corresponding question per passage.
The rubric definitions and question wording are reproduced verbatim, with placeholders for the query and passage text.
Passages were ordered by the returned expected relevance grades.}
\label{fig:jev-rubric-prompt}
\end{figure}

\section{Supervised Fine-Tuning Configuration}
\label{app:sft-configuration}

This appendix records detailed configuration settings for the fine-tuning procedure in Section~\ref{section:sft_config}.
Baseline runs 1--4 denote the four individual causal rerankers described in Section~\ref{section:sft:reproducibility}.
The soup denotes their parameter-averaged checkpoint, used for the headline results in Section~\ref{section:headline}.

Specifically:

\begin{itemize}[leftmargin=0.5cm]

\item \ourptmodeltt provides the initialization for the causal Gaggle rerankers;

\item \ourbidirptmodeltt provides the initialization for the Gaggle (MNTP) rerankers.

\end{itemize}

\noindent MNTP identifies the bidirectionally adapted backbone, not a separate relevance-training objective.

\paragraph{Data sampling and microbatching.}
\label{app:sft-sampling}
For each training instance, we sampled one positive passage and $K$ hard negatives from the query's available passages.
We chose $K$ as the largest value in $\{7,15,23\}$ supported by its negative pool.
Of the 247,534 baseline instances, 1.93\% used seven negatives, 45.85\% used 15, and 52.21\% used 23, forming groups of 8, 16, or 24 query--passage pairs.
Instances were bucketed by $K$ so that each microbatch contained groups of the same size.
Each group contributed one contrastive loss, computed jointly over its positive and sampled negatives.

Within each bucket, we packed complete query groups into per-GPU microbatches, padding the query--passage sequences to the longest sequence in the microbatch.
Packing was limited by both an estimated padded-token budget and a group-count cap.
Token estimates included the prompt, query and passage lengths, and a 30\% safety margin.
The token budget accounted for GPU memory reserved for model parameters, gradients, and optimizer states, with a cap of 60,000 estimated tokens per GPU.
Baseline runs 1, 3, and 4 used microbatch budgets of 39,312 tokens per GPU; run 2 used 39,347.
Microbatches contained at most 64, 32, or 21 query groups for $K=7$, 15, or 23, respectively, limiting them to approximately 512 query--passage sequences.

Gradient accumulation across microbatches and GPUs targeted an effective batch of 128 query groups per optimizer update.
When a single group exceeded the microbatch budget, gradient caching processed its sequences in smaller chunks while retaining the joint LCE loss over all candidates.
Baseline run 4 retained run 1's packing plan and used gradient caching with execution chunks of up to 16,384 tokens on its H100 GPUs.

\paragraph{Optimization.}
Unless specified otherwise, we fine-tuned all retained parameters for one epoch using AdamW~\citep{loshchilov2019adamw} with a learning rate of $5\times10^{-5}$, $\beta=(0.9,0.95)$, weight decay of 0.01, and gradient clipping at a global norm of 12.
The learning rate increased linearly for 150 steps and then followed a cosine decay to zero at the final update.
Baseline run 1 used seed 17 and 1,931 updates.
Baseline training used two GPUs with gradient checkpointing.

We used a common optimization recipe and changed the factors specified by each experiment.
Apart from the diagnostic learning-rate comparison for Gemma and MiniCPM reported in Section~\ref{section:different-backbones}, we did not perform hyperparameter tuning.
The comparisons characterize effectiveness under this common recipe.

\paragraph{Evaluation and numerical precision.}
All SFT comparisons used top-100 BM25 candidates and nDCG@10, with means computed separately over TREC Deep Learning 2019--2023 and the seven BEIR collections (TREC-COVID, TREC-News, Robust04, NFCorpus, SciFact, SCIDOCS, and FiQA).
The four baseline runs were evaluated using the same harness and inputs as the soup checkpoint.
Table~\ref{tab:sft-shared-settings} records training and evaluation precision in one place.

\paragraph{Hardware and execution.}
Baseline runs 1--3 used two NVIDIA RTX PRO 6000 Blackwell Server Edition GPUs (96 GB each), while run 4 used two H100 GPUs (80 GB each).
Runs 1, 3, and 4 completed 1,931 updates; run 2 completed 1,933.
The self-filtering, self-distillation, causal QPQ, and standard MNTP variants used the same hardware and microbatch budget as baseline run 1.
Both self-filtered variants completed 1,918 updates.
The RLHN-680K, Tevatron MS MARCO, mean-pooling, and mask-appended variants used two H200 NVL GPUs (141 GB each) with budgets of 60,000 tokens per GPU.
The RLHN-680K and Tevatron runs completed 5,060 and 3,112 updates, respectively; the prompt and pooling variants completed 1,928--1,931 updates.
The nanochat-d34 run used four RTX A6000 GPUs (48 GB each), a budget of 11,121 tokens per GPU, and 1,929 updates.
The $5{\times}10^{-5}$ Gemma-4-E2B and Qwen3.5-4B-Base runs used the baseline microbatch budget and completed 1,931 updates, using two RTX PRO GPUs and two H200 NVL GPUs, respectively.
The $1{\times}10^{-5}$ MiniCPM run used the same budget and completed 1,931 updates, on two RTX PRO GPUs for the first 1,683 updates and two H200 NVL GPUs for the rest.
The $1{\times}10^{-5}$ Gemma-4-E2B run used two H200 NVL GPUs, the baseline microbatch budget executed in chunks of at most 8,192 tokens per GPU, and completed 1,931 updates.
The pointwise CE run used two NVIDIA RTX PRO 6000 Blackwell GPUs (96\,GB each) and completed 1,931 updates, with microbatches of eight query--passage pairs per GPU.

\paragraph{Readout and checkpoint selection.}
The default readout used the last non-padded token; the mask-appended prompt used the token immediately before the mask, and mean pooling averaged final hidden states over all non-padding tokens.
The final checkpoint was used for each baseline evaluation; checkpoint-saving intervals were 200 updates for runs 1--2, 50 for run 3, and 100 for run 4.
The soup was produced by parameter averaging without further optimization.

\begin{table*}[t]
\startsftrows
\centering
\tableformat
\setlength{\tabcolsep}{1.5pt}

\begin{tabular}{@{}rll@{}}
\toprule
& Setting & Value \\
\midrule
\rowcolor{black!5}
\sftrowlabel{row:shared:tuning} & Fine-tuning & All retained parameters; no frozen layers or LoRA \\
\sftrowlabel{row:shared:scoring} & Scoring parameters & Two pre-trained LM output rows: 4,352 parameters; no new head \\
\rowcolor{black!5}
\sftrowlabel{row:shared:implementation} & Implementation & Native backbone; scaled dot-product attention (SDPA) \\
\sftrowlabel{row:shared:precision} & Training precision & FP32 weights, gradients, and optimizer states; TF32 matrix multiplication \\
\rowcolor{black!5}
\sftrowlabel{row:shared:evalprecision} & Evaluation precision & BF16 backbone; FP32 scoring rows and scoring computation; final hidden states cast to FP32 \\
\sftrowlabel{row:shared:checkpointing} & Gradient checkpointing & Enabled \\
\rowcolor{black!5}
\sftrowlabel{row:shared:lengths} & Query / passage limits & 128 / 512 tokens \\
\sftrowlabel{row:shared:negatives} & Negatives per group & $K\in\{7,15,23\}$ \\
\rowcolor{black!5}
\sftrowlabel{row:shared:batch} & Target effective batch & 128 query groups \\
\sftrowlabel{row:shared:groupcap} & Microbatch group cap & \mytt{max\_groups}=32, scaled by group size \\
\rowcolor{black!5}
\sftrowlabel{row:shared:memory} & Batch-memory estimate & 1 GiB per 1,000 estimated tokens \\
\sftrowlabel{row:shared:optimizer} & Optimizer & AdamW, $\beta=(0.9,0.95)$ \\
\rowcolor{black!5}
\sftrowlabel{row:shared:lr} & Learning rate & $5\times10^{-5}$ \\
\sftrowlabel{row:shared:weightdecay} & Weight decay & 0.01 \\
\rowcolor{black!5}
\sftrowlabel{row:shared:clipping} & Gradient clipping & Global norm 12 \\
\sftrowlabel{row:shared:schedule} & Schedule & 150 warm-up updates, then cosine decay to zero \\
\rowcolor{black!5}
\sftrowlabel{row:shared:epochs} & Training epochs & 1 \\
\sftrowlabel{row:shared:temperature} & LCE temperature & $\tau=1$ \\
\rowcolor{black!5}
\sftrowlabel{row:shared:distributed} & Distributed training & Two GPUs with distributed data parallelism \\
\sftrowlabel{row:shared:workers} & Data workers & 6 \\
\rowcolor{black!5}
\sftrowlabel{row:shared:saving} & Checkpoint interval & 200 updates \\
\bottomrule
\end{tabular}
\caption{Default settings for supervised fine-tuning and evaluating individual Gaggle rerankers.
Run-specific exceptions are described in Appendix~\ref{app:sft-configuration}.
}

\label{tab:sft-shared-settings}
\end{table*}

\end{document}